\documentclass[twocolumn,preprintnumbers,floatfix,superscriptaddress,nofootinbib]{revtex4}
\usepackage[utf8]{inputenc}
\usepackage{setspace}
\usepackage{booktabs}
\usepackage{rotating}
\usepackage{url,comment}
\usepackage{times}
\usepackage{latexsym}
\usepackage{graphicx, graphics, amssymb, slashed, xcolor,amsthm, array}
\usepackage{subfigure}
\usepackage{listings} 
\usepackage{afterpage}
\usepackage{physics}
\usepackage[colorlinks=true,linkcolor=blue,citecolor=blue,urlcolor=blue]{hyperref}
\usepackage{here}
\usepackage{MnSymbol} 
\usepackage{braket}
\usepackage{ulem}

\def\old#1{\iffalse {#1} \fi}
\def\new#1{{#1}}
\def\comment#1{\iffalse {#1} \fi}
\def\sout#1{\iffalse {#1} \fi}

\begin{document}

\vspace*{-10mm}

\title{A Time-Varying Fine Structure Constant from Naturally Ultralight 
Dark Matter}


\author{Dawid Brzeminski}
\email{dbrzemin@umd.edu}
\affiliation{Maryland Center for Fundamental Physics, Department of Physics, University of Maryland, College Park, MD 20742, U.S.A.}

\author{Zackaria Chacko}
\email{zchacko@umd.edu}
\affiliation{Maryland Center for Fundamental Physics, Department of Physics, University of Maryland, College Park, MD 20742, U.S.A.}

\author{Abhish Dev}
\email{adev@umd.edu}
\affiliation{Maryland Center for Fundamental Physics, Department of Physics, University of Maryland, College Park, MD 20742, U.S.A.}

\author{Anson Hook}
\email{hook@umd.edu}
\affiliation{Maryland Center for Fundamental Physics, Department of Physics, University of Maryland, College Park, MD 20742, U.S.A.}


\vspace*{1cm}

\begin{abstract} 

We present a class of models in which the coupling of the photon to an 
ultralight scalar field that has a time-dependent vacuum expectation 
value causes the fine structure constant to oscillate in time. The 
scalar field is assumed to constitute all or part of the observed dark 
matter. Its mass is protected against radiative corrections by a 
discrete $\mathbb{Z}_N$ exchange symmetry that relates the Standard Model to 
several copies to itself. The abundance of dark matter is set by the 
misalignment mechanism. We show that the oscillations in the fine 
structure constant are large enough to be observed in current and 
near-future experiments.

\end{abstract}

\maketitle

\section{Introduction}

With the discovery of the Higgs boson~\cite{Aad:2012tfa,Chatrchyan:2012ufa}, the Standard Model (SM) of 
particle physics is now complete. However, cosmological observations 
tell us that visible matter constitutes only about 20\% of the matter in 
the universe, the remainder being composed of some form of non-luminous 
dark matter~\cite{Aghanim:2018eyx}. Despite contributing five times more to the energy budget 
of the universe than visible matter, the basic nature of dark matter has 
thus far eluded us.

Moduli are among the most well-motivated and compelling dark matter 
candidates~\cite{Goldberger:1999uk,Arvanitaki:2009fg,Chacko:2012sy,Coradeschi:2013gda,Kane:2015jia}. They often arise in ultraviolet-complete theories such as 
string theory, where their vacuum expectation values play a role in 
determining the values of fundamental parameters such as the fine 
structure constant. The properties of moduli make them ideal candidates 
to play the role of dark matter. During inflation, quantum fluctuations 
are stretched to super horizon length scales, with the result that all 
scalars lighter than the Hubble scale acquire large random vacuum 
expectation values. After inflation ends, a light scalar such as a 
modulus remains frozen away from its minimum by Hubble friction. At late 
times, the modulus begins to 
oscillate about its minimum, contributing to the energy density of the 
universe as a component of dark matter. This framework, termed the 
misalignment mechanism~\cite{Linde_1987}, is a natural way of 
generating an abundance of extremely cold bosonic dark matter from 
light fields such as moduli.

There has recently been a surge of interest in finding new ways to 
search for very light dark matter, see 
e.g.~\cite{Damour2010_convention,Damour:2010rm,Damour:2012rc,Khmelnitsky:2013lxt,Stadnik:2014tta,Arvanitaki:2015iga,Graham:2015ouw,Graham:2015ifn,Stadnik:2015xbn,Delaunay:2016brc,Berlin:2016woy, 
Geraci:2016fva,Krnjaic:2017zlz,Roberts:2017hla,Arvanitaki:2017nhi,DeRocco:2018jwe,Geraci:2018fax,Irastorza:2018dyq,Carney:2019cio,Guo:2019ker,Grote:2019uvn,Dev:2020kgz,Stadnik:2020bfk}. 
One of the the most exciting approaches to finding modulus dark matter 
is applicable when it is so light that its Compton wavelength is 
macroscopic, so that the period of its oscillations, set by the inverse 
of its mass, can be directly observed. As moduli set the value of 
fundamental constants, they can be searched for by looking for time 
dependence ("chronovariance") of these parameters. As a wave of modulus 
dark matter washes over us, it oscillates with a characteristic $\cos 
\omega t \approx \cos m t$ time dependence, where $m$ is the mass of the 
modulus field. Therefore a natural way to search for ultralight modulus 
dark matter is to look for time dependence of the fundamental constants. 
For early work on time variation of fundamental parameters, see for 
example 
\cite{Dirac:1937ti,Chodos:1979vk,Terazawa:1981ga,Bekenstein:1982eu, 
Marciano:1983wy}.

For dark matter to be observable in this way, it must be strongly 
coupled enough to be detected over noise while oscillating at a low 
enough frequency that its effects are not averaged away. The condition 
that the frequency be low translates into the requirement that the mass 
be small. Since the mass of the modulus receives radiative corrections 
that depend on the size of its couplings, there is some tension between 
the condition that the mass remain small and the requirement that the 
couplings are large enough to be observable, resulting in a naturalness 
problem. 

To understand the naturalness problem, we parametrize the couplings of the 
modulus $\phi$ to the electromagnetic field in the form,
 \begin{equation}
-\frac{1}{4 e^2} F^{\mu \nu} F_{\mu \nu}
\left( 1 - d_e \kappa \phi \right)
\label{modulusCoupling}
 \end{equation}
 where $\kappa$ is defined as $\sqrt{4 \pi G_N}$, where $G_N$ is Newton's
 constant and $d_e$ is a dimensionless coupling constant that 
parametrizes the strength of the interaction. For a given mass of the 
modulus $\phi$ there is a bound on the amplitude of its oscillations, 
and therefore on the magnitude of the variation of $\alpha$, from the 
condition that the modulus contribute no more to the energy budget of 
the universe than the observed dark matter contribution. Current 
experiments are most sensitive to a modulus mass in the range 
$10^{-22}-10^{-5}$ eV. The corresponding bound on $d_e$ stands at $d_e 
\lesssim 10^{-8} - 10^{-1}$, with the exact value depending on the 
modulus mass. This coupling in Eq.~(\ref{modulusCoupling}) gives rise to 
a radiative contribution to the mass of the modulus at two 
loops~\footnote{In the absence of other couplings, the interaction in 
Eq.~(\ref{modulusCoupling}) can be redefined away.  When the electron is 
present, this redefinition results in a correction to the electron 
coupling to the photon.  In this basis, the natural expectation is that 
the divergence appears at two-loops.}, 
 \begin{equation} \label{Eq: naturalness}
\delta m^2 \sim \frac{e^2 d_e^2 \kappa^2}{\left(16 \pi^2 \right)^2} \Lambda^4 \; 
 \end{equation}
 where $\Lambda$ represents an ultraviolet cutoff.  As an example, 
consider a modulus mass of $10^{-18}$ eV for which the bound stands at $d_e 
\lesssim 10^{-3}$.  For this value of $d_e$ there is a contribution to 
its mass coming from loops involving the top quark of order $10^{-11}$ 
eV. Thus, there must be a cancellation in the mass squared of this 
scalar between the top quark contribution and the bare mass to at least 
one part in $10^{14}$. More generally, for any value of the modulus mass 
there is an upper bound on $d_e$ above which the conflict comes to the 
fore. Furthermore, any understanding of the abundance of the modulus 
requires a consistent ultraviolet completion, since the behavior of a 
modulus in the early universe is very complicated and is extremely 
sensitive to the presence of other fields.

In this paper, we present a framework in which the modulus that controls 
the fine structure constant can naturally remain ultralight while 
constituting all of the observed dark matter in the universe. In order 
to make the modulus naturally light, we employ the mechanism proposed in 
Ref.~\cite{Anson1802}.  Accordingly, we introduce $N$ copies of the SM, 
where $N$ is a number of order a few. If the same modulus controls the 
fundamental parameters of all the $N$ copies of the SM while nonlinearly 
realizing the $\mathbb{Z}_N$ symmetry, then the sum of the contributions 
to the potential of the modulus from each of the copies of the SM 
cancels to a very high degree of accuracy.  The end result is then a 
modulus with a parametrically smaller mass than naively expected.

A very attractive feature of this framework is that, for certain ranges 
of the modulus mass and couplings, the misalignment mechanism naturally 
allows the modulus to constitute all of the observed dark matter. For a 
given mass and couplings of the modulus, we can determine the thermal 
corrections to its potential in the early universe.  The simplest region 
of parameter space is that in which the Hubble friction holds the 
modulus in place and oscillations only begin after thermal effects have 
become subdominant to the zero temperature potential.  Even in this most 
simple of scenarios, there exists a vast region of parameter space in 
which the modulus can constitute all of dark matter. In general, finite 
temperature effects can drastically alter the behavior of the modulus at 
early times.  They can act either to reduce or increase the abundance of 
dark matter by relaxing the modulus to the minimum or maximum of its 
potential at high temperatures.  Including these effects expands the 
region of parameter space in which the modulus can play the role of dark 
matter.

Prospects for probing the time variation of fundamental constants due to 
ultralight scalars are very promising 
\cite{Arvanitaki2015,Arvanitaki:2015iga,Arvanitaki2018,Arvanitaki:2017nhi}. 
In fact, existing data from experiments 
\cite{VanTilburg2015,Hess2016,Berge2017_MICROSCOPE,Hoyle1999_EP_CuPb,Schlamminger2008_EP_BeTi,Baggio2005_AURIGA,Kennedy:2020bac,Vermeulen:2021epa} 
is already sufficient to place constraints on the scenario we propose in 
the mass range $10^{-22}\text{eV} <m_{\phi} < 10^{-5}\text{eV}$. These 
limits can be significantly improved in the future by increasing the 
integration times in the optical-optical clocks and ultimately by taking 
advantage of the anticipated $^{229}$Th nuclear-optical clock 
\cite{Arvanitaki2015,nuclearclock}. Further advancements can be achieved 
with the recently proposed earth and space based atomic gravitational wave 
detectors, which will rely on atom interferometry 
\cite{Arvanitaki2018,Coleman_MAGIS100,Badurina2019_AION,Bertoldi2019_AEDGE}. 
The MAGIS and AION experiments \cite{Coleman_MAGIS100,Badurina2019_AION} 
will be earth based interferometers that are planned to gradually 
increase in size to finally reach a length of 1 km. MAGIS is currently 
building a 100 m interferometer \cite{Coleman_MAGIS100}, which should be 
able to probe the proposed model in the $10^{-16}\text{eV} <m_{\phi} < 
10^{-14}\text{eV}$ mass range. The 1 km stage should increase this range 
to $10^{-16}\text{eV} \lesssim m_{\phi} \lesssim 10^{-12}\text{eV}$. 
Once built, the AION experiment will be able to improve the bounds set 
by MAGIS at both the 100 m and 1 km scale for the same range of masses 
$m_{\phi}$. The AEDGE experiment \cite{Bertoldi2019_AEDGE} is a proposed 
continuation of the AION experiment that will take advantage of 
satellites in order to increase the scale of the detector to thousands 
of kilometers. While it is a very distant prospect, it will have greatly 
improved sensitivity for masses in the range, $10^{-19}\text{eV} 
<m_{\phi} < 10^{-13}\text{eV}$.

The outline of this paper is as follows. In Sec.~\ref{Sec: Model}, we 
discuss the framework and explain the mechanism that protects the mass 
of the ultralight modulus. In Sec.~\ref{Sec: history}, we study the 
cosmological history of this class of models and show that the 
misalignment mechanism can allow the modulus to constitute all of the 
observed cold dark matter. In Sec.~\ref{sec: signal} we present our 
results. We determine the current bounds on this class of models and 
outline the region of parameter space that will be explored by current 
and future experiments. We conclude in Sec.~\ref{Sec: conclusion}

\section{The Framework} \label{Sec: Model}

In this section we construct a class of models in which the fine 
structure constant oscillates in time, and which are free of naturalness 
problems. We consider a complex scalar $\Phi$ that is charged under an 
approximate ${\it global}$ U(1) symmetry. This field is assumed to 
acquire a vacuum expectation value, spontaneously breaking the global 
symmetry at a scale denoted by $f$. It is convenient to employ an 
exponential parametrization of $\Phi$,
 \begin{equation}
    \Phi = \left( \frac{f + \rho}{\sqrt{2}} \right) \exp{\frac{i \varphi}{f}}.
 \end{equation}
 Here $\rho$ represents the radial mode in the potential for the scalar 
field after symmetry breaking, while $\varphi$ denotes the 
pseudo-Nambu-Goldstone boson. We write $\varphi = \varphi_0 + \phi$, 
where $\varphi_0$ is the vacuum expectation value (VEV)
of $\varphi$ while $\phi$ represents the 
fluctuation about this expectation value. The field $\phi$
is assumed to couple to 
the electromagnetic field strength as shown in 
Eq.~(\ref{modulusCoupling}). Although this interaction is 
nonrenormalizable, it can be generated by coupling $\Phi$ to some heavy 
charged fermions that are integrated out. As a consequence of this 
coupling, changes in the background value of $\varphi$ will cause 
variations in the fine structure constant. However, this coupling 
violates the U(1) global symmetry explicitly. We therefore expect that, 
in general, it will generate a potential for $\varphi$, leading to a 
quadratically divergent mass.  Since the parameter $d_e$ controls both 
the amplitude of the modulation as well as the magnitude of the 
potential, we require some mechanism that protects the potential against 
large radiative corrections while still admitting an observable signal. 
As we now explain, this can be done by employing a discrete $\mathbb{Z}_N$ 
symmetry under which the pseudo-Goldstone $\varphi$ transforms 
nonlinearly~\cite{Anson1802}.
 
 We introduce $N$ copies of the SM, each with its own matter content and 
gauge groups. We label each of these different copies by an index $i$, 
where $i$ runs from 1 to $N$. The $N$ copies of the SM are related by a 
discrete $\mathbb{Z}_N$ symmetry under which $i \rightarrow (i + 1)$. To each 
copy of the SM we add a heavy Dirac fermion $\Psi_{i}$ that carries unit 
charge under the corresponding $U(1)_{Y}$ hypercharge gauge symmetry, 
but not under any of the other gauge groups. Each of the $N$ fermions 
$\Psi_i$ is assumed to have a Yukawa coupling to the scalar $\Phi$. 
Under the $\mathbb{Z}_N$ symmetry these fields transform as,
 \begin{equation}
    \Psi_{k} \to \Psi_{k+1}, \, F_{k} \to F_{k+1}, 
    \, \Phi \to \Phi \exp{i \frac{2 \pi}{N}}.
 \end{equation}
 The discrete symmetry forces the Yukawa couplings, gauge couplings and 
masses to be same across all the $N$ sectors. The Lagrangian for 
the $\Psi_i$ is restricted to have the form,
 \begin{equation}
 \mathcal{L}  \supset \sum_{k}^{N} \left\lbrace \bar{\Psi}_{k} i \slashed{D}_{ k}\Psi_{k} -  (M - y \Phi e^{i \frac{2 \pi k}{N}}-y \Phi^{\dag} e^{-i \frac{2 \pi k}{N}}) \bar{\Psi}_{k} \Psi_{k} \right\rbrace .
 \end{equation}
 Here the Yukawa coupling $y$ can be chosen to be real without loss of 
generality as any phase of complex $y$ can be absorbed into the 
definition of the complex scalar $\Phi$.

\begin{figure}
    \centering
    \includegraphics[width = 0.4\textwidth]{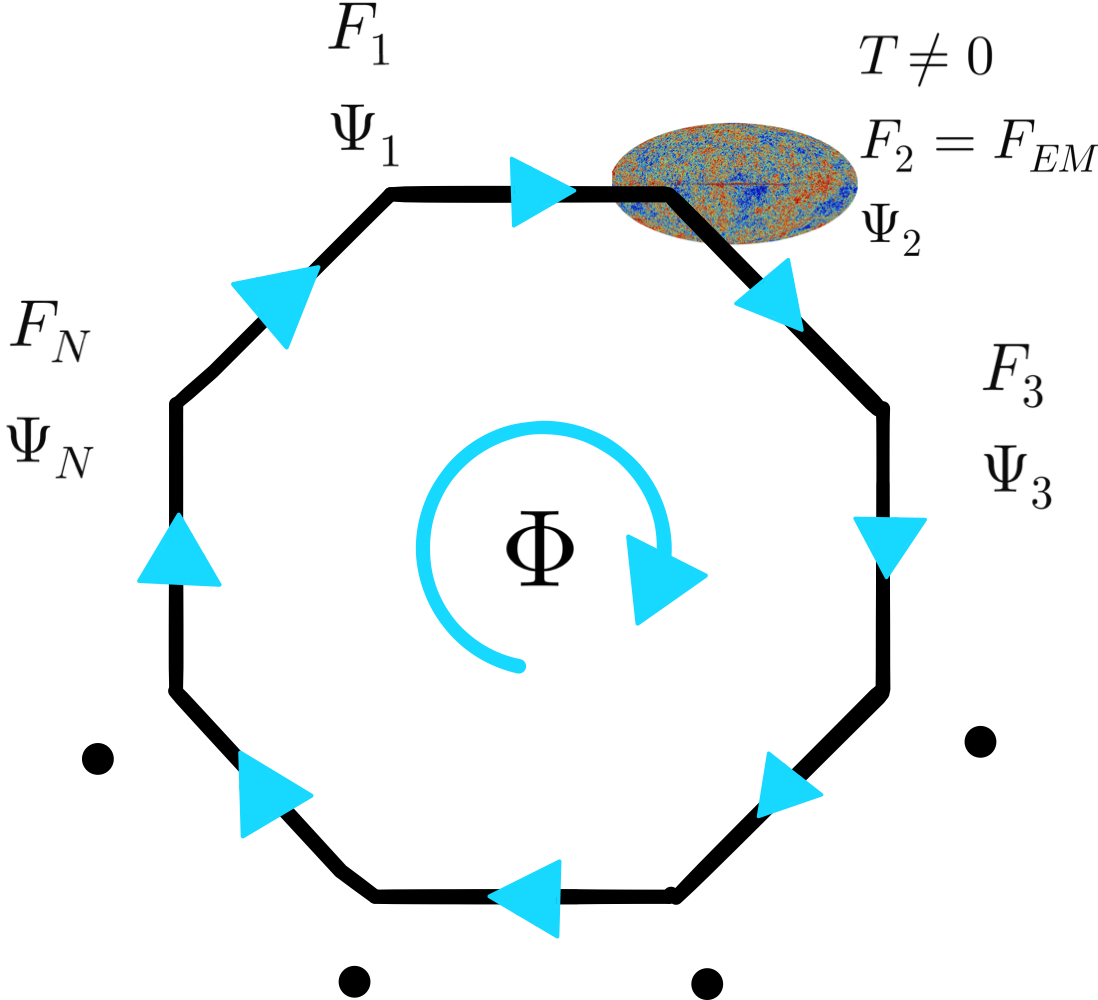}
    \caption{A cartoon picture of the $\mathbb{Z}_N$ model.  There are $N$ copies of the Standard Model and $N$ copies of a heavy hypercharged vector-like fermion $\Psi$ that are cyclically exchanged under the $\mathbb{Z}_N$ symmetry.  A complex scalar $\Phi$ transforms linearly under the $\mathbb{Z}_N$ and has a Yukawa coupling to the fermions $\Psi_k$.}
    \label{fig:LS}
\end{figure}
 After symmetry breaking, the relevant part of the Lagrangian becomes
 \begin{equation}
\mathcal{L} \supset \sum_{k}^{N} \bar{\Psi}_{k}( i \slashed{D}_{ k} - M_{k})\Psi_{k} ,
 \end{equation}
 where the effective mass of each fermion depends on the VEV of the 
pseudo-Nambu Goldstone boson $\varphi$
 \begin{equation} \label{eq. vector-like mass}
M_{k} = M \left(1 - \epsilon \cos \left( \frac{\varphi}{f} + \frac{2 \pi k}{N}\right) \right) \;.
 \end{equation}
 Here the parameter $\epsilon$ is defined as $\epsilon \equiv 
{\sqrt{2}yf}/{M}$. After integrating out the fermions $\Psi_{k}$ we get 
a contribution at one loop to the effective potential for the 
pseudo-Nambu Goldstone boson,
 \begin{equation} 
 \label{phi_potential_full}
V(\varphi) = -\frac{1}{8 \pi^{2}} \sum^{N}_{j=1} \int\limits^{\Lambda^2}_{0}  d k^{2}_{E} k^{2}_{E} \log \left[ k^{2}_{E} + M^{2}_{j} \right],
 \end{equation}
 where $\Lambda$ is an ultraviolet cutoff. This integral can be 
evaluated exactly.

 After dropping terms that do not depend on $\varphi$ and terms that 
vanish in the limit $M/\Lambda \rightarrow 0$, we obtain
 \begin{eqnarray}\label{V}\nonumber
    V(\varphi)&=&\frac{1}{16 \pi^2}\sum_{j=1}^N\left[ - 2 \Lambda^2 M^2 (1-\epsilon_j)^2 + 2 M^4 (1-\epsilon_j)^4 \log \frac{\Lambda}{M}\right.\\
  &&+  \left. \frac{M^4}{2}(1-\epsilon_j)^4 -2 M^4 (1-\epsilon_j)^4 \log (1- \epsilon_j) \right],
 \end{eqnarray}
 where $\epsilon_j\equiv \epsilon \cos(\varphi/f + 2\pi j/N)$. The first two terms 
in Eq.~(\ref{V}) are quadratically divergent and logarithmically 
divergent. However, they can be seen to independent of $\varphi$ 
as long as $N>4$ by virtue 
of the following identity~\cite{Anson1802}
 \begin{equation}\label{identity}
    \sum^{N}_{j=1} \cos^{m} \left( \frac{\varphi}{f}+\frac{2 \pi j}{N} \right) = \left\{ \begin{matrix}
    0&& m= \text{ odd} < N\\
    \\
    \frac{N}{2^{m}} \binom{m}{m/2} && m= \text{ even} <N \\
    \\
    \frac{N}{2^{N-1}} \cos \left( \frac{N \varphi}{f} \right)+C && m= N
    \end{matrix} \right.,
 \end{equation} 
where $C$ is a constant.
 The finite contribution is, however, $\varphi$-dependent. It can be 
extracted by expanding the last term in Eq.~(\ref{V}) in the small 
parameter $\epsilon$. To all orders in $\epsilon$, we get contributions 
from each sector that only differ by the phase of the cosine. Due to the 
identity in Eq.~(\ref{identity}), all the terms up to 
$\mathcal{O}(\epsilon^{N-1})$ cancel leaving the leading 
$\varphi-$dependence to appear at $\mathcal{O}(\epsilon^{N})$. These 
cancellations are a direct consequence of the unbroken $\mathbb{Z}_N$ symmetry. 
After performing the integral in \eqref{phi_potential_full}, the leading 
$\varphi$-dependent piece arises at order $\epsilon^{N}$
 \begin{equation} 
 \label{eq: phi_potential zero temp}
V(\varphi) = \frac{M^{4} \epsilon^{N}}{8 \pi^{2}} F(N) \cos \left(\frac{N \varphi}{f}\right)\left[1 +\mathcal{O}(\epsilon)\right],
 \end{equation}
where $F(N)$ is given by
 \begin{equation}
 F(N) = 2^{1-N} N \sum_{l=0}^{4}  \binom{4}{l} \frac{(-1)^{l}}{N-l}.
 \end{equation}

 This potential has $N$ minima at the locations $\varphi_m/f=(2 m-1)\pi/N$ 
where $m$ runs from 1 to $N$. Integrating out the fermions $\Psi_i$ also 
leads to an effective coupling of $\varphi$ to the gauge kinetic terms of 
each of the $N$ hypercharge gauge fields at low energies,
 \begin{equation}
    \mathcal{L}_{eff} \supset -\sum^{N}_{j=1} \frac{\epsilon}{24 \pi^{2}} \cos \left( \frac{\varphi}{f} + \frac{2 \pi j}{N} \right)F^{2}_{j}.
 \end{equation}
 One of these $N$ copies corresponds to the hypercharge gauge boson of 
the SM.  Since the effective potential Eq.~\eqref{eq: phi_potential zero temp} is 
invariant under $\varphi \to \varphi + 2 \pi f /N$, we have freedom to shift 
$\varphi$ so that the coupling to SM hypercharge gauge boson simplifies to
 \begin{equation} 
 \label{eq: B coupling}
    \mathcal{L}_{\varphi BB} = - \frac{\epsilon}{24 \pi^{2}} \cos \left( \frac{\varphi}{f} \right)F^{2}_{\text{Y}}.
 \end{equation}
Since in our normalization $B_\mu = A_\mu - W^3_\mu$, the coupling to SM photons after 
spontaneous symmetry breaking takes the same form,
  \begin{equation} 
 \label{eq: EM coupling}
    \mathcal{L}_{\varphi \gamma \gamma} = - \frac{\epsilon}{24 \pi^{2}} \cos \left( \frac{\varphi}{f} \right)F^{2}_{\text{EM}}.
 \end{equation}
 When $\varphi$ performs small oscillations about one of the minima 
$\varphi_m$ of \eqref{eq: phi_potential zero temp}, it gives rise to a 
linear term in the coupling to the photon,
 \begin{equation} \label{eq: linear coupling}
    \mathcal{L}_{\varphi \gamma \gamma} = \frac{\epsilon}{24 \pi^{2} f} \sin \left( \frac{\varphi_{m}}{f} \right) \phi F^{2}_{\text{EM}},
 \end{equation}
 where $\varphi_{m}$ is the $m$-th minimum of the potential, 
Eq.~\eqref{eq: phi_potential zero temp}, and $\varphi = \varphi_{m} + 
\phi$. Comparing to Eq.~(\ref{modulusCoupling}), we see that in terms of 
the parameters of our model, $d_e$ is given by,
 \begin{equation}
\label{Eq: de}
    d_{e} = \frac{2 \epsilon \alpha \abs{\sin \left ( \varphi_m/f \right )}}{3 \pi  \kappa f} .
 \end{equation}
 We are now at a stage where we can demonstrate the improvement in 
naturalness that our model affords. As discussed in the Introduction, 
the natural expectation is that the mass of $\phi$ scales as
 \begin{equation} 
 m^2 \sim \frac{e^2 d_e^2 \kappa^2}{\left(16 \pi^2 \right)^2} M^4 \; ,
 \end{equation}
 which is just Eq.~(\ref{Eq: naturalness}) with the divergence cut off by 
the mass of the new charged fermion, in this case $M$. To see how our 
model compares to this, we expand Eq.~(\ref{eq: phi_potential zero temp}) 
about the minimum to obtain
 \begin{eqnarray} 
\label{eq: mass term stuff}
m^2_\phi &=& \frac{M^4 N^2 \epsilon^N}{8 \pi^2 f^2} F(N) \\ \nonumber 
         &\approx& \frac{e^2 d_e^2 \kappa^2}{\left(16 \pi^2 \right)^2} M^4 \left ( \frac{1152 \pi^6 N^2 F(N) \epsilon^{N-2}}{e^6 \sin^2 \left( \frac{\varphi_{m}}{f} \right)}    \right ) . 
 \end{eqnarray}
 The term in brackets represents the improvement with respect to the 
naive estimate. In order for this expression to be valid, we had assumed 
that $N>4$.  Because this correction factor is proportional to 
$\epsilon^{N-2}$, we see that as long as $\epsilon$ is small the mass is 
parametrically smaller than expected from naive considerations.  This 
demonstrates that our construction can indeed solve the naturalness 
problem discussed in the Introduction. We note that the mass term 
for the modulus is generated at one-loop order rather than being given 
by the naive 2-loop estimate in Eq.~(\ref{Eq: naturalness}). This is 
because both the mass of the modulus and its coupling to the hypercharge 
gauge boson arise at the same one-loop order when the heavy fermions 
$\Psi_i$ are integrated out, rather than the mass being generated 
radiatively from the coupling.

At this stage, the new fermions $\Psi_k$ are electrically charged stable 
fermions.  If the reheat temperature is larger than their mass, then 
these particles obtain a thermal abundance and will tend to overclose 
the universe if their masses lie above the TeV scale. In order to avoid 
this, we allow each of the $N$ $\Psi_k$ to decay by introducing a small 
mixing with the right-handed $\tau$ lepton of the corresponding sector 
through the interaction
 \begin{equation}
    \mathcal{L} \supset \sum_k^N m \Psi_k \tau^c_k .
\end{equation}
For $M= 10$ TeV, we require $m \gtrsim 100$ eV in order to have the $\Psi$ 
particles decay prior to Big Bang nucleosynthesis (BBN) and not pose 
a cosmological problem.

\section{Cosmological history} \label{Sec: history}

In this section, we will consider the cosmological history of this class 
of models. We work under the assumption that of the $N$ sectors, ours is 
the only one that is reheated after inflation\footnote{One way such a 
scenario can arise is if there are $N$ separate inflatons, one for each 
sector. Each of the $N$ inflatons is assumed to reheat only its own 
sector. Then the inflaton that slow rolls to its minimum last is 
associated with the sector that is identified with the SM, while the 
other $(N - 1)$ sectors are not reheated.}. We further assume that 
$\varphi$ is homogenized as a result of inflation. The scalar field 
evolution is governed by the equation
 \begin{equation} 
 \label{eq: EOM}
\ddot{\varphi}+3 H \dot{\varphi}+\frac{\partial V}{\partial \varphi} = 0 \;,
 \end{equation}
where the potential is given by 
 \begin{equation} 
 \label{eq: phi potential general}
V = V_0(\varphi) + V_T(\varphi,T) \;.
 \end{equation}
 Here $V_0$ is the zero temperature potential given in Eq.~(\ref{eq: 
phi_potential zero temp}) while $V_T(\varphi,T)$ is the contribution to 
the potential from finite temperature effects. An explicit expression 
for $V_T$ may be found in Appendix~\ref{Appendix: Thermal calculations}.  
\new{At temperatures $T \gg M$, the finite temperature contribution to 
the potential can be approximated as,
 \begin{equation}
 \label{eq: phi potential high temp}
 V_T(\varphi,T)
\approx \frac{\epsilon}{6} T^2 M^2 \cos \left( \frac{\varphi}{f} \right) 
\, .
 \end{equation}
 Once the temperature falls below its mass, the vector-like fermion 
$\Psi$ begins to exit the bath. Accordingly the form of the thermal 
contribution to the potential undergoes a change. In the temperature 
range $M \gtrsim T \gtrsim 100$ GeV it is well approximated by the 
expression,
 \begin{equation} 
 \label{eq: phi potential mid temp}
 V_T(\varphi,T) \approx \left(\frac{61\epsilon \alpha^{2} q^{2}_{F}  }{216 \cos^{4}\theta_{W}} T^{4}+\frac{4\epsilon M^{\frac{5}{2}} T^{\frac{3}{2}}}{(2\pi)^{\frac{3}{2}}} e^{-\frac{M}{T}}\right)\cos \left( \frac{\varphi}{f} \right)  \, .
 \end{equation}
For $T\lesssim M/20$, the exponential suppression of the second term in 
the bracket means that it can be neglected.}

 \new{The evolution of the scalar field $\varphi$ at early times is 
governed by the extent to which its oscillations are damped by Hubble 
friction. To keep track of whether the system is underdamped or 
overdamped at temperature $T$, we define the parameter
 \begin{equation} 
 \label{eq: m/H general}
   \eta(T) \equiv \frac{4 m^{2}(T)}{H^{2}(T)}  , 
 \end{equation}
 where $m(T)$ represents the contribution to the mass of the modulus 
from finite temperature effects,
 \begin{equation} \label{eq: m(T) definition}
    V_{T}(\varphi,T) \equiv m^2(T) f^2 \cos \left(\frac{\varphi}{f}\right) \, .
 \end{equation}
 For $\eta(T) < 1$ the system is overdamped at temperature $T$, while 
for $\eta(T) > 1$ it is underdamped.}

\new{At early times $\eta(T)$ increases as the universe cools 
down. This can easily be seen from Eq.~(\ref{eq: phi potential high 
temp}) by noting that when $T\gg M$, the mass scales linearly with 
temperature, $m(T)\propto T$, while the Hubble parameter decreases 
faster, $H\propto T^2$. Eventually, at temperatures $T$ of order $M$, 
the growth of $\eta(T)$ slows down, reaching its maximal value $\eta_p$ 
at a temperature $T_{p} = \frac{2M}{5}$. The value of $\eta_p$ is given 
by
 \begin{equation}
    \eta_{p} \equiv \eta(T_{p}) = 
\frac{2250\sqrt{5}M_{pl}^2}{ e^{\frac{5}{2}} 
\pi^{\frac{7}{2}} g_{*}}\frac{\epsilon }{f^{2}} \;.
 \end{equation}
  Here $M_{pl}$ is the reduced Planck mass and $g_{*} = 106.75$ is the 
effective number of degrees of freedom at $T\lesssim M$, consisting 
of just the SM fields. 
 For $\eta_{p} \gtrsim 1$, the field oscillates about the minimum of the 
thermal potential at temperatures $T$ of order $M$. The oscillations 
begin at a temperature $T_{osc}$ such that $m(T_{osc}) = H(T_{osc})$, 
given by
 \begin{equation}
    T_{osc} = \sqrt{\frac{15\epsilon M_{pl}^2}{ \pi^{2} g^{\Psi}_{*} f^2}}M \; .
 \end{equation}
 Here $g^{\Psi}_{*} = 110.25$ is the effective number of degrees of 
freedom at temperatures $T \gg M$, which consists of the SM fields and 
the associated fermion $\Psi$.}

 \new{Below the temperature $T_p$, $\eta(T)$ decreases rapidly as the 
exponential suppression of the second term in Eq.~\eqref{eq: phi 
potential mid temp} takes effect. Eventually at temperatures $T\lesssim 
M/20$, the contribution from the first term in Eq.~\eqref{eq: phi 
potential mid temp} becomes dominant. In this temperature regime, the 
mass of the modulus and the Hubble parameter scale identically, 
$m(T)\propto H(T) \propto T^2$. As a result $\eta(T)$ reaches a terminal
value $\eta$, given by
 \begin{equation} 
 \label{eq: m/H}
   \eta \equiv \eta(T\lesssim M/20) = \frac{305 \alpha^{2} M_{pl}^{2}}{3\pi^{2}g_{*} \cos^{4} \theta_{W}} \frac{\epsilon}{f^{2}} . 
 \end{equation}
 The parameters $\eta_p$ and $\eta$ play an important role in the 
description of the system. The value of $\eta_{p}$ indicates whether the 
system undergoes oscillations when $T\gtrsim M$, while $\eta$ determines 
the behavior of the system at temperatures $T\lesssim M/20$. These 
parameters are related as
 \begin{equation}
     \eta_{p} = \frac{1350 \sqrt{5} \cos^4 \theta_{W}}{61\alpha^2 e^{\frac{5}{2}}\pi^{\frac{3}{2}}} \eta \approx 8300 \eta \, .
 \end{equation}
 From this relation we can see that in the regime which is always 
overdamped at early times, $\eta \ll 10^{-4}$, the modulus field is 
effectively frozen until the temperature-independent part of the 
potential becomes dominant. In contrast, for $\eta > 1$ the 
oscillations of the field begin at the temperature $T_{osc} \gtrsim M$ 
and continue till the present time. In the intermediate regime $10^{-4} 
\lesssim \eta \lesssim 1$, the field oscillates at temperatures $T$ of 
order $M$. However, these oscillations have ceased by the time the 
temperature falls below $M/20$ and only resume once the 
temperature-independent contribution to the potential begins to 
dominate.}

 \new{In what follows below we obtain analytic expressions for the 
contribution of the modulus $\phi$ to the energy density of the 
universe. We focus on the limiting cases of $\eta \ll 10^{-4}$ and $\eta 
\gtrsim 1$, deferring the intermediate range of $\eta$ to our numerical 
study.}

 \new{\subsection{$\eta \ll 10^{-4}$ } \label{Section: history overdamped}}

For \new{$\eta \ll 10^{-4}$} the early time 
evolution is especially simple as Hubble friction freezes the field in 
place at high temperatures, so that its evolution is independent of 
$V_T$. It follows that for sufficiently small $\eta$ the field 
is effectively \new{fixed} until the 
mass approaches its zero temperature value, $m=m_{\phi}$. Oscillations 
begin when $3H\sim m_{\phi}$, corresponding to a temperature
 \begin{equation} \label{eq: Ts}
    T_{s}= \left(\frac{10}{\pi^2 g_{*}}\right)^{1/4} \sqrt{M_{pl} m_{\phi}} \;.
 \end{equation}
At this point, we make the 
standard approximation that $\varphi$ transitions instantly from an 
overdamped harmonic oscillator to an underdamped one. At this stage the 
potential Eq.~\eqref{eq: phi potential general} is dominated by the zero 
temperature term,
 \begin{equation} \label{eq: phi_potential_cold}
\begin{aligned}
V(\varphi) &= \frac{M^{4} \epsilon^{N}}{8 \pi^{2}} F(N) \cos \left(\frac{N \varphi}{f}\right)\\
&\equiv \frac{m^{2}_{\phi} f^{2}}{N^{2}} \cos \left(\frac{N \varphi}{f}\right).
\end{aligned}
 \end{equation}
 The mass $m_{\phi}$ is constant so the field will oscillate as an 
underdamped harmonic oscillator around a minimum $\varphi_{m}$, with the 
amplitude decaying as
 \begin{equation}
    \phi \propto a(t)^{-3/2} .
 \end{equation}
 Since the field is frozen, the value of $\varphi$ when $3H\sim 
m_{\phi}$ can lie anywhere in the range $\varphi_{s} \in (0,2\pi f)$. 
Therefore $\varphi$ can fall into any of the $N$ minima $\varphi_{m}$ of 
the zero temperature potential, Eq.~\eqref{eq: phi_potential_cold}, with 
equal probability. Consequently the initial misalignment value of the 
excitation around that minimum $\varphi_{m}$ can lie anywhere in the 
range $\phi_{s} \in (-\pi f/N,\pi f/N)$. Due to this randomness in the 
initial condition and the homogeneity of $\phi$, it is not possible to 
determine the exact value of the field today. However, what can be done 
instead is to average over all possible initial misalignment values, 
$\phi_{s} \in (-\pi f/N,\pi f/N)$, given that all are equally likely. 
The expectation value of the energy density $\bar{\rho}_\phi$ can be 
related to an effective initial amplitude of the field 
$\phi_{s,\text{eff}}$, where $\phi_{s,\text{eff}} = 
\sqrt{\braket{\phi^{2}_{s}}}$. The value of 
$\sqrt{\braket{\phi^{2}_{s}}}$ can be obtained by averaging over all 
values of the initial misalignment, leading to $\phi_{s,\text{eff}} = 
\frac{\pi f}{\sqrt{3}N}$. As a result the expected amplitude of the 
field today is given by
 \begin{equation} 
 \label{amplitude}
   \phi_{0} \approx \phi_{s,\text{eff}} \left( \frac{T_{0}}{T_{s}} \right)^{3/2} =  \frac{\pi f}{\sqrt{3} N} \left( \frac{T_{0}}{T_{s}} \right)^{3/2}, 
 \end{equation}
 where $T_0$ represents the current temperature of the Universe. 
This leads to a final result for the energy density in $\phi$ today, 
 \begin{equation} \label{Eq: overdamped rho}
    \bar{\rho}_{\phi}^{o} = \frac{1}{2}m^{2}_{\phi} \phi_{0}^{2} \approx \frac{\pi^2 f^2 m_{\phi}^2}{6 N^2} \left( \frac{T_{0}}{T_{s}} \right)^{3} \propto f^2 m_{\phi}^{1/2}.
 \end{equation}

\subsection{$\eta \gtrsim 1$}

\begin{figure*}[t] 
    \centering
    \includegraphics[width=0.495\linewidth]{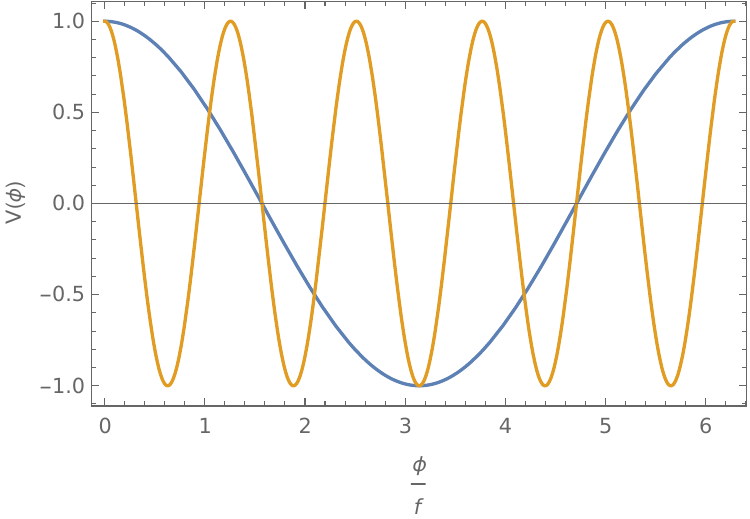}
    \includegraphics[width=0.495\linewidth]{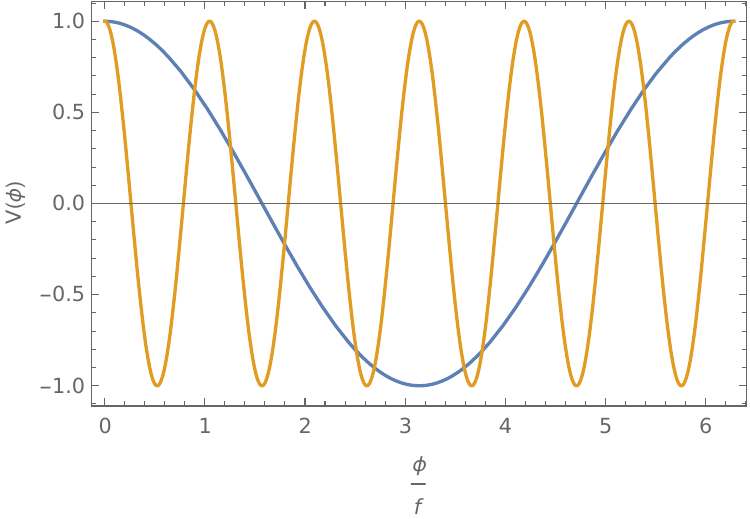}
    \caption{Example of thermal (blue) and zero temperature (orange) potentials for $N=5$ (left) and $N=6$ (right) }\label{fig. potentials}
\end{figure*}

In this subsection, we consider the range of parameter space in which 
the behaviour of $\varphi$ in the early universe \new{at temperatures 
$T \lesssim M/20$} is like that of a harmonic oscillator 
that is either underdamped or close to critically damped.  When 
discussing finite temperature effects, it is necessary to distinguish 
between the cases when $N$ is even and $N$ is odd.  The reason for this 
can be seen from Fig.~\ref{fig. potentials}. In the case of even $N$, 
the thermal potential pushes $\varphi$ towards a maximum of the zero 
temperature potential. At late times, this results in an increase in the 
amplitude of oscillations, leading to enhanced abundance of dark matter 
and a larger signal. For odd $N$, the thermal potential pushes $\varphi$ 
towards a minimum of the zero temperature potential, decreasing the 
amplitude of oscillations at late times resulting in a suppressed 
abundance. In this case the situation is even worse because the minimum 
that $\varphi$ is pushed towards is the one in which there is no linear 
coupling to the photon, so that the signal under consideration is 
greatly suppressed.  Rather than a linear coupling, this minimum leads 
to a quadratic coupling, which gives rise to different 
phenomenology~\cite{Stadnik:2015kia,Sibiryakov:2020eir} that we do not 
consider here.

\paragraph{{\bf $N$ odd}} \label{sec: history odd N}

In the early universe $\varphi$ is driven towards the minimum of the 
finite temperature potential, $\varphi_m = \pi f$.
 From Fig.~\ref{fig. potentials} we see that the minimum of the finite 
temperature potential is also a minimum of the zero temperature 
potential. Consequently the range of values of $\varphi$ at the time 
when oscillations begin, $H = 3 m_\phi$, is significantly smaller than 
initial $(0, 2 \pi f)$.

It follows from this that for each $m_{\phi}$, there is a minimal value 
of $\eta$ above which this new range covers only the central minimum of 
the zero temperature potential, $\varphi_{s} \in ((\pi-\pi/N) f, 
(\pi+\pi/N) f)$. Above this critical value of $\eta$, which we denote by 
$\eta_0$, $\varphi$ always ends up in the central minimum. Furthermore, 
from Eq.~\eqref{eq: linear coupling} we see that at this central minimum, 
$\varphi_m = \pi f$, the linear coupling 
of $\phi$ to photons vanishes, resulting 
in a greatly suppressed signal. 
 \new{The value of $\eta_0$ can be well approximated by considering 
evolution only in the regime $T\lesssim M/20$. The evolution at higher 
temperatures is less efficient at focusing the range of values of 
$\varphi$ due to the sudden drop of the modulus mass $m(T)$ at $T\sim M$, 
which has the effect of regenerating the amplitude of the field. 
The subsequent evolution of $\varphi$ in the regime when 
$T\lesssim M/20$ and $\eta<1$ scales with temperature as
 \begin{equation} \label{eq: early evolution}
    \varphi -\pi f \propto T^{\left(1- \sqrt{1-\eta } \right)/2} .
 \end{equation}
It follows that $\eta_0$ satisfies the condition,
 \begin{equation} \label{eq: no signal condition}
    \left( \frac{T_{s}}{M/20} \right)^{\left(1- \sqrt{1-\eta_0 } \right)/2} = \frac{1}{N}.
 \end{equation}
 From this we obtain an expression for $\eta_0$,
 \begin{equation} \label{eq: eta0 value}
     \eta_0 = 1-\left(1-\frac{2 \ln N}{\ln\frac{M}{20 T_{s}}}\right)^2 \, .
 \end{equation}
}

 We see that in the case of odd $N$, for $\eta \geq \eta_0$, the field is 
pushed to the minimum in which $\phi$ does not have the coupling shown 
in Eq.~(\ref{modulusCoupling}), resulting in a greatly suppressed 
signal. The abundance of $\phi$ is also suppressed limiting its 
contribution to dark matter.

\paragraph{{\bf $N$ even}}

For even values of $N$ the picture is completely different. The minimum 
of the finite temperature contribution to the potential, which is at 
$\varphi_m = \pi f$, now coincides with the maximum of the zero 
temperature potential, as can be seen in Fig. \ref{fig. potentials}. 
Therefore, in the \new{regime} 
$\eta \gtrsim 1$ the 
initial value of the modulus $\varphi$ at the time when oscillations 
begin will be close to $\pi$ and the initial amplitude will be very 
close to
 \begin{equation} \label{eq: initial amplitude lightly overdamped}
    \phi_{s} = \frac{\pi f}{N} ,
 \end{equation}
 This is very different from the case when the system is overdamped. 

The amplitude of the oscillations of $\phi$ today depends on how close 
the field is to the minimum of the finite temperature potential at the 
time that the oscillations begin. 
 \new{Given that the mass of the scalar changes as a function of 
temperature, the simplest way to determine the behavior of $\varphi(T)$ 
at these early times is to employ the conservation of the number density of 
$\varphi$.  The comoving number density of $\varphi$ is approximately 
conserved in the limit that $\varphi$ is close enough to a minimum that 
its potential is quadratic and as long as the WKB approximation holds, 
$dm/dt \ll m^2$.  The conserved comoving number density is given by
 \begin{equation} \label{eq: number density}
    n = a^3 m(T) \left(\varphi-\pi f\right)^2 .
 \end{equation}
 Based on these considerations the effective initial value of the 
modulus $\phi_{s,eff}$ when the zero temperature potential begins to 
dominate can be obtained as
 \begin{equation} 
 \label{eq: phi initial value underdamped}
\abs{\phi_{s,\text{eff}}-\frac{\pi f}{N}} = \abs{\varphi_{s,\text{eff}}-\pi f} = \frac{\pi f}{\sqrt{3}} \sqrt{\frac{m(T_{osc})}{m_{\phi}}} \left( \frac{T_s}{T_{osc}} \right)^{\frac{3}{2}} \;.
 \end{equation} 
 We continue to track the evolution of the field after it rolls down to 
the new minimum, which happens at temperatures $T$ of order $T_s$.}

In contrast to the case when \new{$\eta \ll 10^{-4}$}, the position of $\varphi$ 
when the zero temperature contribution to the potential begins to 
dominate is very close to $\varphi = \pi f$. Since this point 
corresponds to an extremum of both the finite temperature and zero 
temperature contributions to the potential, the gradient of the 
potential at $\varphi = \pi f$ vanishes independent of the temperature. 
The fact that the potential in the neighborhood of this point is very 
flat leads to a delayed onset of 
oscillations~\cite{Lyth_1992,Arvanitaki_anharmonic} and interesting 
phenomenological signatures~\cite{Arvanitaki:2019rax,Huang:2020etx}. As 
shown in Appendix \ref{Appendix: amplitude underdamped}, for the 
parameters considered in this paper this delay is long enough to ensure 
that when the oscillations about the new minimum begin, the 
contributions to the potential from finite temperature effects are 
already negligible. Therefore, to a good approximation we can assume 
that the potential has the zero temperature form in Eq.~\eqref{eq: 
phi_potential zero temp} from the time that the oscillations begin.

It is tempting to assume that oscillations about the true minimum occur 
in a harmonic potential, starting from the temperature $T_s$, defined in Eq.~\eqref{eq: Ts}, 
with an initial amplitude $\phi_{s,\text{eff}} = \frac{\pi f}{N}$ and 
continuing till today. This leads to the following expression for the 
contribution of the oscillating field to the energy density today,
 \begin{equation} \label{Eq: underdamped rho naive}
    \bar{\rho}^{\text{naive}}_\phi \approx \frac{\pi^2 f^2 m_{\phi}^2}{2 N^2} \left( \frac{T_{0}}{T_{s}} \right)^{3}.
 \end{equation}
 However, this expression does not give the correct result for the energy 
density as it fails to account for the the delay in the time at which 
the oscillations begin. To include this effect we employ the following 
empirical approximation \cite{Arvanitaki_anharmonic},
\new{\begin{equation} \label{Eq: underdamped rho}
    \bar{\rho}_{\phi}^{u} = C\left(\sqrt{\frac{m_{\phi}}{m(T_{osc})}} \left( \frac{T_{osc}}{T_{s}} \right)^{\frac{3}{2}}\right) f^2 m_{\phi}^2 \left( \frac{T_{0}}{T_{s}} \right)^{3} \propto f^2 m_{\phi}^{1/2}.
\end{equation}
}
Here
 \begin{equation} \label{eq: c(y)}
    C(y) = \frac{0.23}{N^{2}} \left( \beta(y) +4 \ln \beta(y) \right)^2
 \end{equation}
 and
 \new{\begin{equation}
    \beta(y) = \ln y + \ln  \frac{2^{1/4} \sqrt{3}}{\pi^{1/2} \Gamma(5/4) N } \; .
 \end{equation}
 }
 The coefficient $C(y)$ defined in Eq.~\eqref{eq: c(y)} corresponds to 
the correction arising from the delayed onset of oscillations.

The difference between the energy density in Eq.~\eqref{Eq: underdamped 
rho} and the naive estimate in Eq.~\eqref{Eq: underdamped rho naive} 
turns out to be quite significant. 
 \new{For $M = 10$ TeV the correction factor is 
$\bar{\rho}_{\phi}^{u}/\bar{\rho}^{\text{naive}}_\phi \approx 9$ for 
$N=6$ and $\bar{\rho}_{\phi}^{u}/\bar{\rho}^{\text{naive}}_\phi \approx 
10$ for $N=10$.} \comment{updated}

 Since the field is pushed very close to the central maximum of 
$V_0(\varphi)$ potential, one could imagine that instead of the field 
homogeneously rolling down to a unique minimum, quantum 
fluctuations would push the field in some regions of space to the 
minimum on the other side of the hill, resulting in domain walls. 
However, just after
inflation, quantum fluctuations are given by~\cite{Lyth_1992},
 \begin{equation}
    \sqrt{\braket{(\delta \varphi_{i})^2}} \simeq \frac{H_{inf}}{2 \pi}. 
 \end{equation}
 The ratio of fluctuations to the average value of the field is 
constrained by CMB observations, 
 \begin{equation}
    \frac{\sqrt{\braket{(\delta \varphi_{i})^2}}}{\varphi_{i,eff} - \pi f} \sim \frac{H_{inf}}{f} \lesssim 10^{-5} .
 \end{equation}
 Because this quantity changes at most logarithmically between the end 
of inflation and now~\cite{Visinelli:2009zm} and the emergence of domain 
walls requires $\delta \varphi_{s} \sim \varphi_{s,eff} - \pi f $, this 
scenario does not take place in our model.

\section{Signal} \label{sec: signal}

In this section, we determine the region of parameter space populated by 
our model and explore the implications for experiment. Following the 
conventions employed in 
\cite{Damour2010_convention,Damour:2010rm,Arvanitaki2015}, we 
parametrize the changes to $\alpha$ in terms of the variable $d_{e}$ 
defined in Eq.~\eqref{modulusCoupling},
 \begin{equation} 
 \label{Eq: experiment}
    d_{e} = \frac{\Delta \alpha}{\alpha} \frac{1}{\kappa \phi_{local}} = \frac{\Delta \alpha}{\alpha} \frac{m_\phi }{\kappa \sqrt{2 r \rho_{DM}^{local}}} \;.
 \end{equation}
 Here $\Delta \alpha$ represents the amplitude of oscillations of the 
fine structure constant while $\phi_{local}$ denotes the amplitude of 
oscillations of the scalar field at the location of the earth. We 
introduce a parameter $r$ that represents the fractional contribution of 
$\phi$ to the total energy density in dark matter,
 \begin{equation}
   \rho_{\phi}^{local}= \frac{1}{2} m_\phi^2 \phi_{local}^2 =  r \rho_{DM}^{local}.
 \end{equation}
 Experiments place a bound on $\Delta \alpha/\alpha$ for a given 
frequency of oscillation $m_\phi$. Then, for a given dark matter 
fraction $r$, this can be translated into a bound on $d_e$ using 
Eq.~(\ref{Eq: experiment}).

 As can be seen from Eq.~(\ref{Eq: de}), the value of $d_e$ is a 
function of the parameters $\epsilon$, $f$, $M$ and $N$. From this 
equation we further see that the value of $d_e$ also depends on the 
minimum $\varphi_{m}$ that the field settles in, which is in turn 
determined by the initial conditions. In our analysis, we take 
$\abs{\sin \left ( \varphi_m/f \right )}$ to be the average of available 
values, e.g. for $N=6$ we have $\braket{\abs{\sin \left ( \varphi_m/f 
\right )}}=2/3$ in the regime \new{$\eta \ll 
10^{-4}$} and $\braket{\abs{\sin \left ( \varphi_m/f \right )}}=1/2$ in 
the regime \new{$\eta > 1$}. Additionally for 
odd values of $N$ we ignore the contribution of the central minimum 
($\sin \left ( \varphi_m/f \right ) = 0$) to the average. This allows us 
to determine the expected value of $d_e$ consistent with a given set of 
parameters.

In order to make contact with experiment it is convenient to eliminate 
the parameter $\epsilon$ in favor of $m_{\phi}$. The expression for 
$m_{\phi}$ in terms of the other four variables is given in Eq. 
\eqref{eq: phi_potential_cold}. The parameter space of our model can 
then be described in terms of the four variables $f$, $m_{\phi}$, $M$ 
and $N$. Using the results of Section~\ref{Sec: history}, the 
expectation value of the dark matter abundance can also be expressed in 
terms of these parameters, after averaging over the initial conditions 
for the scalar field $\varphi$. The requirement that, after this 
averaging, the field $\phi$ constitutes all of the observed dark matter 
places a restriction on the allowed parameter space and allows us to fix 
the value of $f$ in terms of the three other variables. Then $d_{e}$ is 
determined in terms of $m_{\phi}$, $M$ and $N$. Taking advantage of the 
results of Section \ref{Sec: history}, we can obtain semi-analytical 
expressions for $d_e$ as a function of these three remaining parameters 
in various regimes. These will prove helpful in illuminating the main 
features of the detailed numerical results, which we will present later.

\subsection{Analytic Results for \new{$\eta \ll 10^{-4}$}}

In the overdamped limit, we can use Eqs.~(\ref{eq: mass term 
stuff}),~(\ref{Eq: overdamped rho}) and~(\ref{Eq: de}) to obtain an 
expression for $d_{e}$ in terms of $m_{\phi}$, $M$ and $N$,
 \begin{equation} 
\label{eq: de scaling overdamped}
    d_{e} = A_{1}(N) \left(\frac{m_{\phi}}{10^{-20} \text{ eV}}\right)^{\frac{N+6}{4N}} \left(\frac{M}{10 \text{ TeV}}\right)^{-\frac{4}{N}}.
 \end{equation}
Here 
 \begin{equation}
\begin{aligned}
   &A_{1}(N) = B_{0} \left( \frac{B_{2}^{N-2} }{B_{1}^{2}} \right)^{\frac{1}{2N}} \frac{\left(10^{-20} \text{ eV}\right)^{\frac{N+6}{4N}}}{ \left(10 \text{ TeV}\right)^{\frac{4}{N}}},\\ \\
   &B_{0} = \frac{2 \alpha \braket{\abs{\sin \varphi_{m}}}}{3 \pi \kappa}, \quad B_{1}  = \frac{N^2 F(N)}{8 \pi^{2}},
\end{aligned}
 \end{equation}
 and
 \begin{equation}
\begin{aligned}
    &B_{2} = \frac{\pi^2}{6 N^2} \left( \frac{\pi^2 g_{*}}{10}\right)^{3/4}\frac{T_{0}^{3}}{\rho_{0} M_{pl}^{3/2}} \nonumber.
    \end{aligned}
 \end{equation}
 For convenience, a few values of $A_{1}(N)$ are given in 
Table~\ref{tab: de coefficients}.  This analytic approximation 
reproduces our detailed numerical results up to an accuracy of around 
$10\%$. \comment{checked}                                                                             
From Eq.~(\ref{eq: de scaling overdamped}) we can see that $d_e$ 
decreases as we raise the mass $M$ of the fermions $\Psi$. Then, by 
setting this mass to the lowest value allowed by experiment, we can 
place an upper bound on $d_e$ as a function of $m_{\phi}$ for any given 
value of $N$.

\new{
\begin{table}
\begin{tabular}{|c|c|c|c|c|}
\hline
N & $A_{1}^{N}$ & $A_{2}^{N}$ & $A_{3}^{N}$ & $\eta_{ref}$ \\
\hline
 $5$ & $1.8\times 10^{-9}$ & $7.5\times 10^{-4}$ & N/A & 0.58\\
 \hline
 $6$ & $3.9\times 10^{-8}$ & N/A & $2.7\times 10^{-3}$ & N/A\\
 \hline
 $10$ & $1.4\times 10^{-5}$ & N/A & $1.0\times 10^{-1}$ & N/A  \\
 \hline
 $11$ & $3.2\times 10^{-5}$ & $3.2 \times 10^{-2}$ & N/A & 0.77\\
 \hline
\end{tabular}
\caption{\new{Scaling coefficients for Eq.~\eqref{eq: de scaling overdamped}, Eq.~\eqref{eq: de odd N} and Eq.~\eqref{eq: de scaling underdamped} evaluated for a few values of N.} \comment{updated}}
\label{tab: de coefficients}
\end{table}}

\subsection{Analytic Results for $\eta \gtrsim 1$}

\new{\begin{table*}[t]
\begin{tabular}{|c|c|c|c|c|c|c|}
\hline
$f$ [GeV] & $y$ & $M$ [TeV] & $N$ & $m_{\phi}$ [eV] & $d_e$ & $\epsilon$ \\
\hline
 $2.0\times 10^{16}$ & $3.2\times 10^{-21}$ & 10 & 5 & $10^{-20}$ & $1.8\times 10^{-9}$ & $9.1\times 10^{-9}$ \\
 \hline
 $4.3\times 10^{14}$ & $2.5\times 10^{-16}$ & 10 & 6 & $10^{-13}$ & $1.2\times 10^{-4}$ & $1.5\times 10^{-5}$ \\
 \hline
 $7.2\times 10^{15}$ & $4.6\times 10^{-16}$ & 10 & 10 & $10^{-17}$ & $2.2\times 10^{-4}$ & $4.7\times 10^{-4}$ \\
 \hline
 $1.4\times 10^{16}$ & $3.7\times 10^{-16}$ & 10 & 11 & $10^{-18}$ & $1.9\times 10^{-4}$ & $7.3\times 10^{-4}$\\
 \hline
\end{tabular}
\caption{Typical values of parameters in our model for a few data points. \comment{checked}}
 \label{Tab: values}
\end{table*}}

As $\eta$ increases, the scalar field eventually enters the 
regime $\eta \gtrsim 1$. In this regime, the 
case of odd $N$ is very different from that of even $N$. For odd values 
of $N$, the region where $\eta \gtrsim 1$ does not give rise to an 
observable signal because the field is trapped in the wrong vacuum as 
discussed in Section \ref{Sec: history}. 
 \new{The boundary of this region is marked by $\eta_{0}$, defined in Eq.~\eqref{eq: eta0 value}.}
 The condition $\eta<\eta_{0}$ translates into a restriction on the 
allowed range of $m_{\phi}$, $M$ and $N$. From Eqs.~(\ref{eq: mass term 
stuff}),~(\ref{eq: m/H}) and~(\ref{Eq: de}) we can determine the value 
of $d_{e}$ at the boundary of this region $\eta = \eta_{0}$ to be
 \new{\begin{equation} 
 \label{eq: de odd N}
    d_{e} = A_{2}(N) \left( \frac{\eta_{0}}{\eta_{ref} } \right)^{\frac{N-2}{2(N-1)}} \left(\frac{m_{\phi}}{10^{-11} \text{ eV}}\right)^{\frac{1}{N-1}} \left(\frac{M}{10 \text{ TeV}}\right)^{-\frac{2}{N-1}} ,
 \end{equation}
 where 
 \begin{equation}
 \begin{aligned}
     &A_{2}(N) = B_{0} B_{1}^{-\frac{1}{2(N-1)}} \left( \frac{\eta_{ref}}{B_{3}} \right)^{\frac{N-2}{2(N-1)}} \frac{\left(10^{-11} \text{ eV}\right)^{\frac{1}{N-1}}}{ \left(10 \text{ TeV}\right)^{\frac{2}{N-1}}} \, , \\
     &B_{3}  = \frac{305 \alpha^2 M^{2}_{pl}}{3\pi^{2}g_{*} \cos^{4} \theta_{W}} \, ,\\
     &\eta_{ref} = \eta_{0}(m_{\phi} = 10^{-11} \text{ eV},M=10 \text{ TeV},N) \, .
\end{aligned}
 \end{equation}}
 The numerical values of $A_{2}(N)$ for a few sample points were given 
in Table~\ref{tab: de coefficients}. The allowed parameter space is 
restricted to values of $d_{e}$ lower than Eq.~\eqref{eq: de odd N}. 
\new{The 
line obtained from Eq.~\eqref{eq: de odd N} is found to reproduce our numerical 
results up to a factor of 2.}

We now turn our attention to the case of even $N$. The expressions for 
the energy density in the case \new{$\eta 
\gtrsim 1$}, Eq.~\eqref{Eq: underdamped rho}, and the 
case \new{$\eta \lesssim 10^{-4}$}, Eq.~\eqref{Eq: 
overdamped rho}, differ only by their proportionality constant.  It 
follows from this that $d_{e}$ scales with $m_{\phi}$ and $M$ in the 
same way in both regimes,
 \begin{equation} 
 \label{eq: de scaling underdamped}
    d_{e} = A_{3}(N) \left(\frac{m_{\phi}}{10^{-11} \text{ eV}}\right)^{\frac{N+6}{4N}} \left(\frac{M}{10 \text{ TeV}}\right)^{\frac{-4}{N}},
 \end{equation}
 where 
 \new{
  \begin{equation}
 \begin{aligned}
     &A_{3}(N) = B_{0} \left( \frac{B_{4}^{N-2} }{B_{2}^{2}} \right)^{\frac{1}{2N}} \frac{\left(10^{-11} \text{ eV}\right)^{\frac{N+6}{4N}}}{ \left(10 \text{ TeV}\right)^{\frac{4}{N}}},\\
     &B_{4} = C(y_0) \left( \frac{\pi^2 g_{*}}{10}\right)^{3/4}\frac{T_{0}^{3}}{\rho_{0} M_{pl}^{3/2}},\\
     &y_0 = \left( \frac{81\pi}{61\sqrt{10}}\right)^{1/4} \sqrt{\frac{\cos^2\theta_{W} g^{5/4}_{*} M}{\alpha g^{\Psi}_{*} \sqrt{m_{1} M_{pl}}}} \, , \\
     &\frac{m_{1}}{1\text{ eV}} =\left( B_{1}^{2} B_2^{- 2(N-1)} B_3^{- 2N } M^8\right)^{\frac{1}{N+3}} \, .
 \end{aligned}
 \end{equation}
 Here $m_1$, obtained from Eqs. \eqref{Eq: overdamped rho}, \eqref{eq: 
mass term stuff} and \eqref{eq: m/H}, corresponds to the mass of the 
modulus for which $\eta = 1$ and $\rho_{\phi} = \rho_{DM}$.
The numerical values of $A_{3}(N)$ are given in Table~\ref{tab: de 
coefficients} for a few reference points. Equation \eqref{eq: de scaling 
underdamped} reproduces the detailed numerical solution up to an 
accuracy of about $20\%$. }\comment{updated}

\subsection{Numerical Results}

The analytic solutions found in the subsection above are valid in the 
limiting cases when the scalar field is either highly overdamped or 
highly underdamped. In order to determine the solution in the region of 
parameter space \new{$10^{-4} \lesssim \eta \lesssim 1$} where the system 
transitions \new{between 
these two regimes}, we find it necessary to solve Eq.~\eqref{eq: EOM} 
numerically. We parametrize the model in terms of the four parameters 
$f, m_\phi, N$ and $M$. As explained earlier, lighter fermion masses are 
associated with larger values of $d_e$.  In our study we therefore 
consider two different values of the fermion mass, $M = 10$ TeV and $M = 
1$ TeV, which are close to the current lower bound from collider 
experiments. In addition, we consider four different values of $N$, the 
odd values $N = 5$ and 11 and the even values $N = 6$ and 10. We then 
scan over $f$ for different values of $m_\phi$. For each $(f, m_\phi)$ 
pair, the field was evolved from $T_{i} = f$ to $T = T_{0}$ starting 
from 1000 random initial conditions. In the case of odd $N$, we discard 
any $(f, m_\phi)$ pair such that, for more than $90\%$ of initial 
conditions, the theory ends up in the vacuum with vanishing signal. For 
each point we obtain the contribution of $\phi$ to the dark matter 
density. We also determine the value of $\braket{\abs{\sin \left ( 
\varphi_m/f \right )}}$, which is obtained by averaging over solutions 
with $\sin \left ( \varphi_m/f \right )\neq0$, and use this to find the 
value of $d_e$ at each point from Eqs.~\eqref{Eq: de} and \eqref{eq: 
mass term stuff}.

\begin{figure*}[t]
    \centering
    \includegraphics[width=0.495\linewidth]{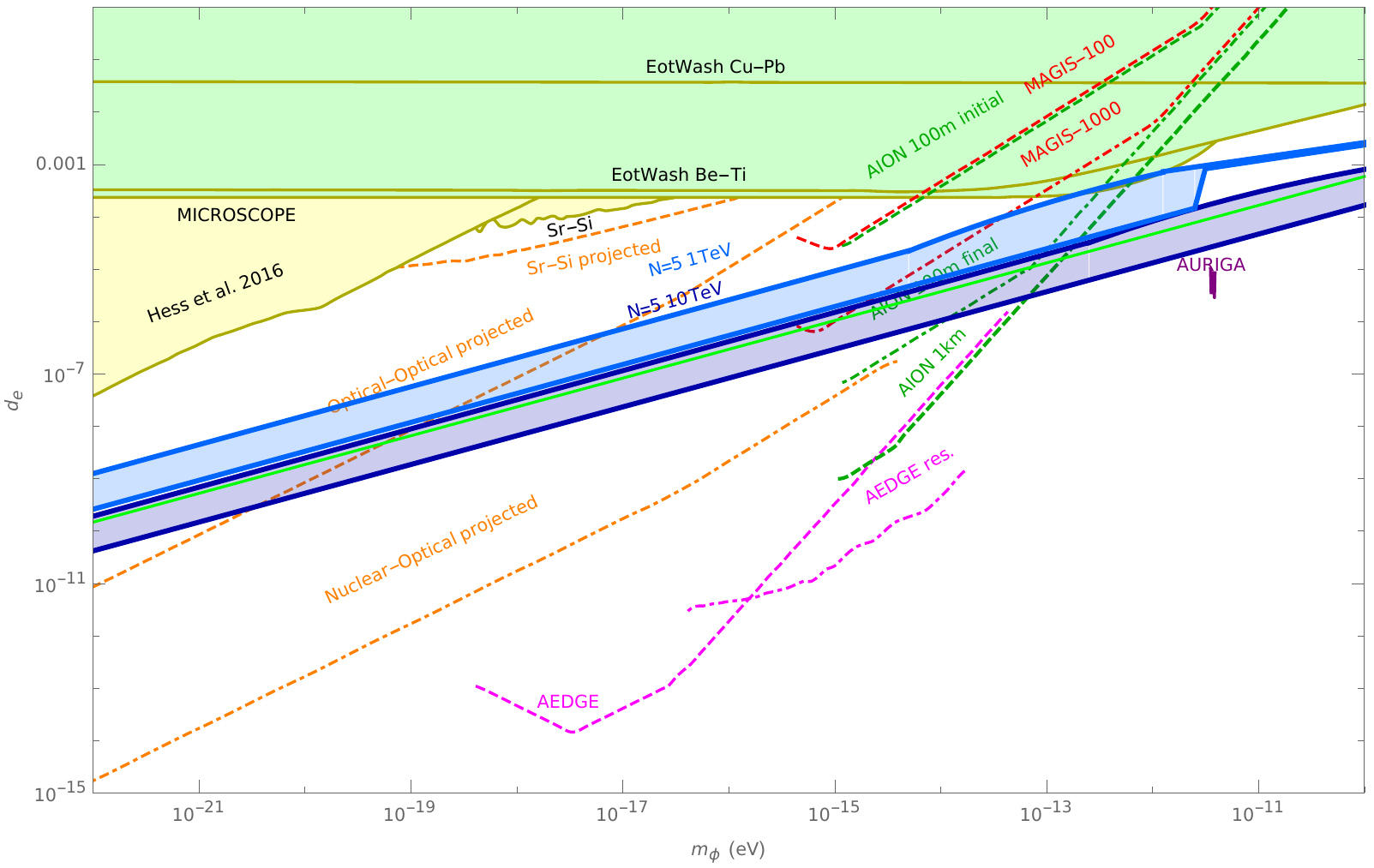}
    \includegraphics[width=0.495\linewidth]{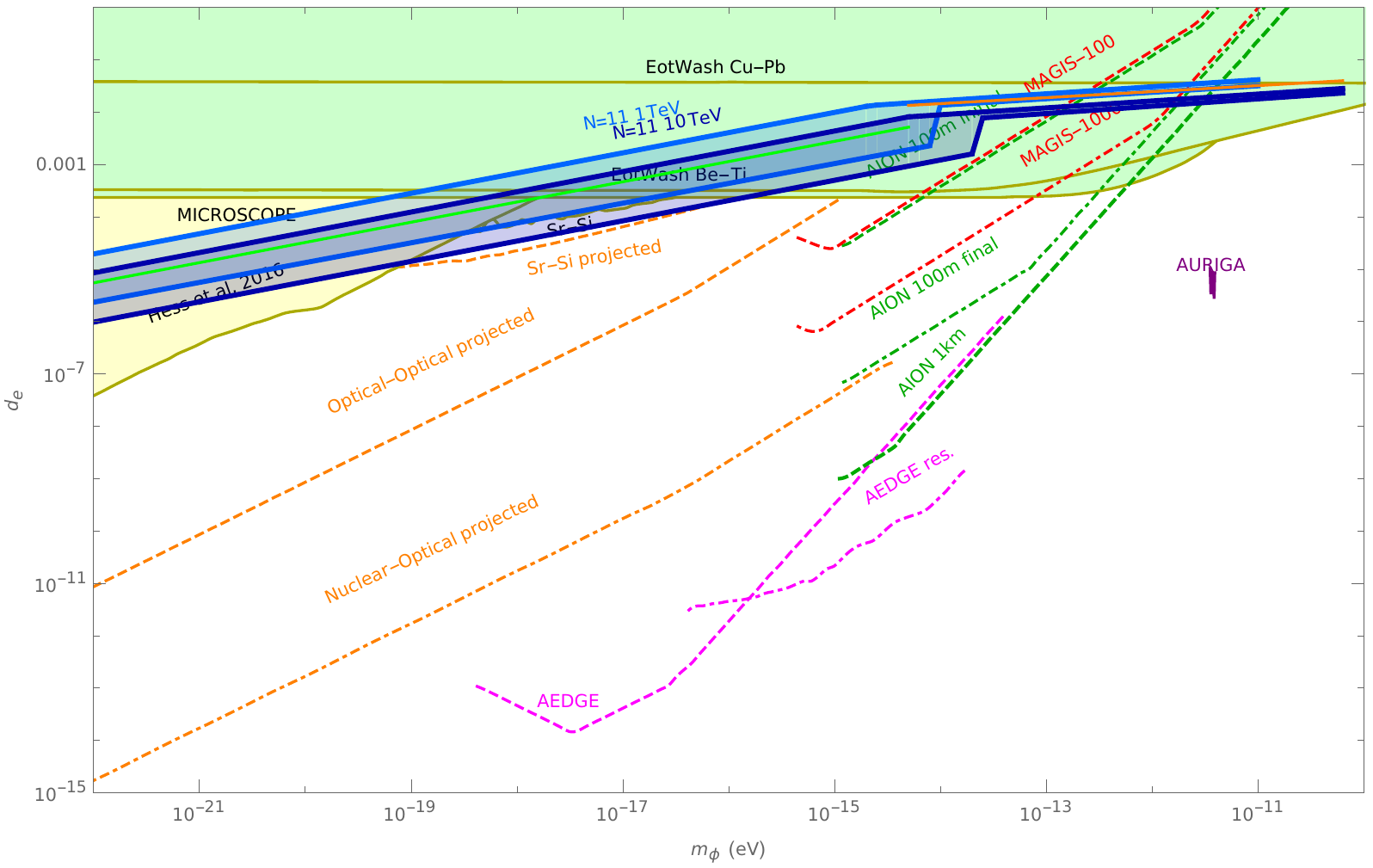}
    \caption{$d_{e}$ vs. $m_{\phi}$ for N=5 (left) and N=11 (right). 
Light ($M=1$ TeV) and dark ($M=10$ TeV) blue bands represent a region 
where no more than $90 \%$ of random initial conditions result in either 
$\rho_{\phi} > \rho_{DM}$ or $\rho_{\phi} < \rho_{DM}$.  The green line 
within the dark blue band is a semi-analytic approximation valid in the 
regime \new{$\eta \ll 10^{-4}$} and the orange 
line (present in the $N=11$ plot only) is a semi-analytic approximation 
of the line above which signal vanishes, both drawn only for $M=10$ TeV. 
Explicit expressions for both lines were given in Eq.~\eqref{eq: de 
scaling overdamped} (green) and Eq.~\eqref{eq: de odd N} (orange). The 
green band gives current constraints from Equivalence Principle 
experiments 
\cite{Berge2017_MICROSCOPE,Hoyle1999_EP_CuPb,Schlamminger2008_EP_BeTi}, 
the yellow band presents the current constraints from atomic clock 
experiments \cite{Hess2016,Kennedy:2020bac} while dashed lines give 
potential reach of the future proposed experiments 
\cite{Bertoldi2019_AEDGE,Badurina2019_AION,Coleman_MAGIS100,Arvanitaki2015}.
}
    \label{n_odd}
\end{figure*}

\begin{figure*}[t]
    \centering
    \includegraphics[width=0.495\linewidth]{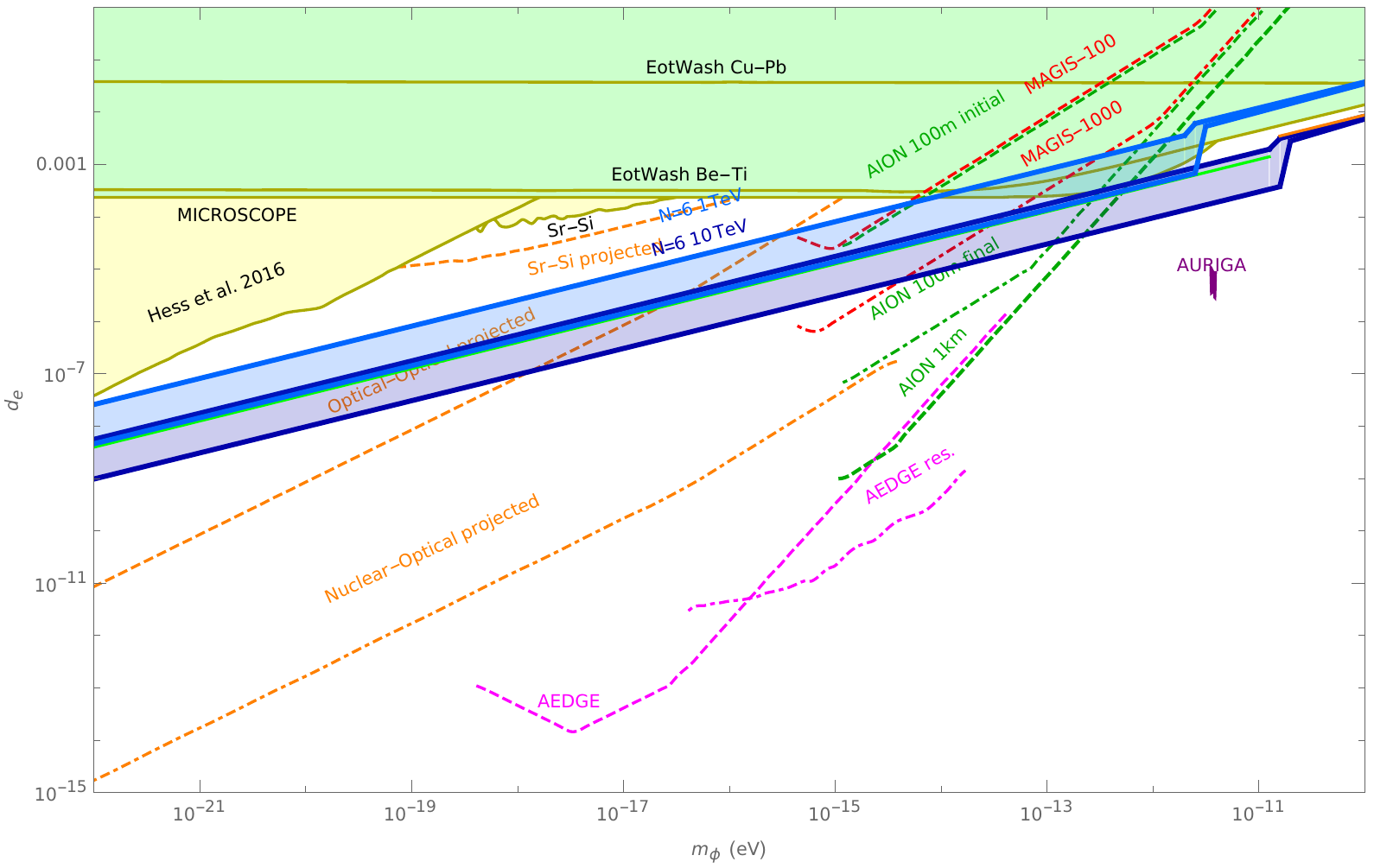}
    \includegraphics[width=0.495\linewidth]{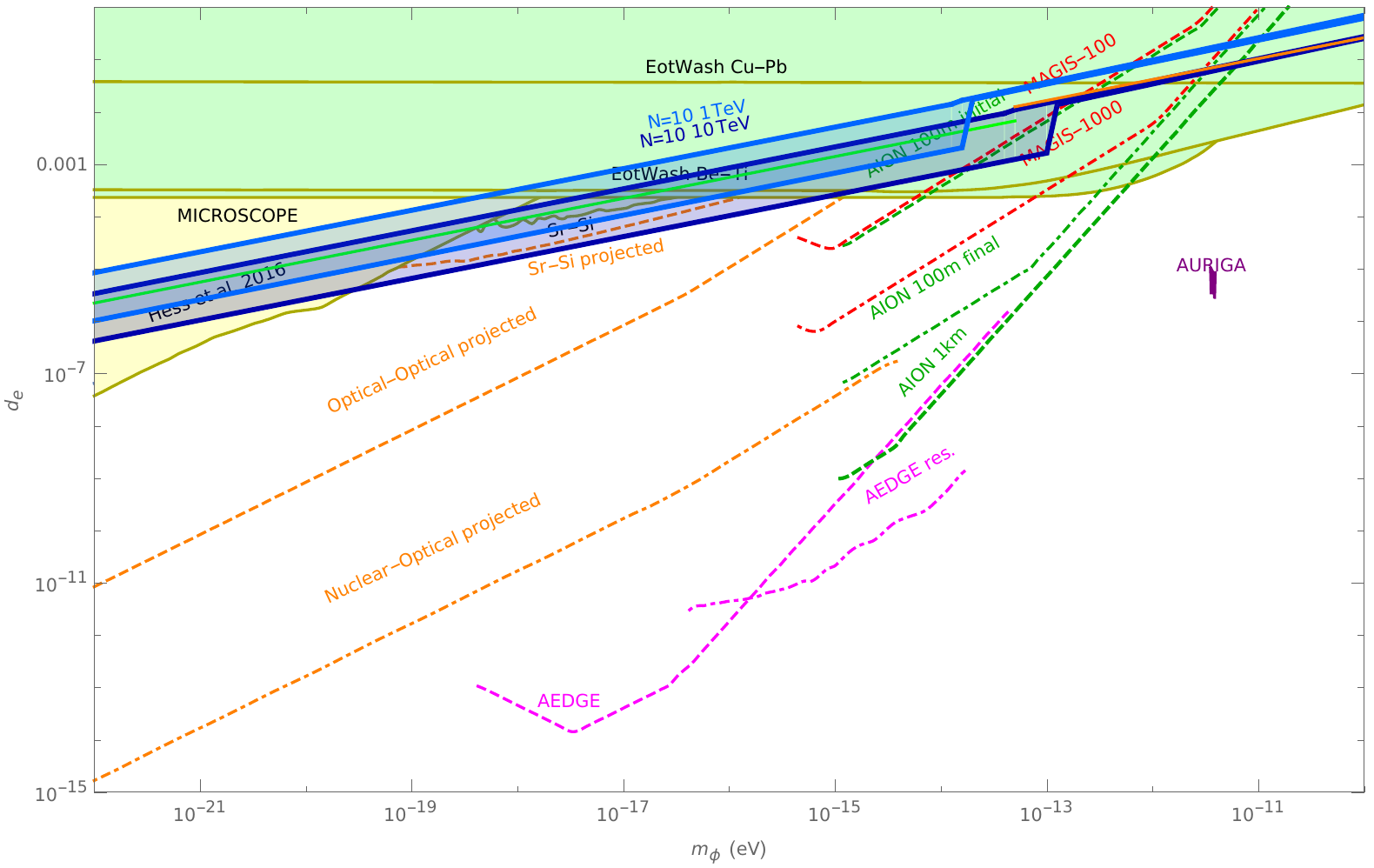}
    \caption{ $d_{e}$ vs. $m_{\phi}$ for N=6 (left) and N=10 (right). 
Light ($M=1$ TeV) and dark ($M=10$ TeV) blue bands represent a region 
where no more than $90 \%$ of random initial conditions result in either 
$\rho_{\phi} > \rho_{DM}$ or $\rho_{\phi} < \rho_{DM}$.  The green line 
within the dark blue band is a semi-analytic approximation valid in the 
regime \new{$\eta \ll 10^{-4}$} 
and the orange line is a semi-analytic approximation in the 
regime \new{$\eta \gtrsim 1$}, 
both drawn only for $M=10$ TeV. Explicit expressions for both lines were 
given in Eq.~\eqref{eq: de scaling overdamped} (green) and 
Eq.~\eqref{eq: de scaling underdamped} (orange). The green band gives 
current constraints from Equivalence Principle experiments 
\cite{Berge2017_MICROSCOPE,Hoyle1999_EP_CuPb,Schlamminger2008_EP_BeTi}, 
the yellow band presents the current constraints from atomic clock 
experiments \cite{Hess2016,Kennedy:2020bac} while dashed lines give 
potential reach of the future proposed experiments 
\cite{Bertoldi2019_AEDGE,Badurina2019_AION,Coleman_MAGIS100,Arvanitaki2015}.
}
    \label{n_even}
\end{figure*}

The results of our numerical study are shown in Figs.~\ref{n_odd} 
and~\ref{n_even}, where we have plotted $d_{e}$ as a function of 
$m_{\phi}$ for these theories, along with the current limits and the 
projected reach of future experiments. Our goal is to identify the 
region of parameter space that is naturally populated by these models. 
To this end, for each $m_{\phi}$ we have singled out the value of 
$d_{e}$ such that, for any $d_{e}$ larger than this, more than $90\%$ of 
points will lead to less than the observed abundance of dark matter, 
$\rho_{\phi} < \rho_{DM}$. Separately, for each $m_{\phi}$ we have 
singled out the value of $d_{e}$ such that, for any $d_{e}$ smaller than 
this, more than $90\%$ of points will lead to more than the observed 
abundance of dark matter, $\rho_{\phi} > \rho_{DM}$. These values have 
been plotted in Figs.~\ref{n_odd} and~\ref{n_even} as the two light blue 
($M = 1$ TeV) lines. The region between these lines, which corresponds 
to the natural parameter space for $M = 1$ TeV, has been shaded in.  The 
natural paramater space for $M = 10$ TeV has also been shown, shaded in 
dark blue.  We explicitly show the values of the parameters for a few 
reference points in Table~\ref{Tab: values}.

In order to simplify the numerical computation, the number of effective 
degrees of freedom in the thermal bath has been held fixed at 
\new{$g_{*} = 106.75$}. The coefficient of the thermal contribution to 
the potential \new{from the first term in Eq.~\eqref{eq: phi 
potential mid temp}}, which also decreases as heavier species go out of 
the bath, has also been held fixed. These simplifications have an effect 
on the temperature at which the final oscillations begin, resulting in 
an overestimate of $d_{e}$ in the case of the lowest values of 
$m_{\phi}$ by a factor close to two. The error is smaller for larger 
values of $m_{\phi}$ since the oscillations begin earlier, and so there 
are a greater number of degrees of freedom in the bath at the 
corresponding $T_s$.

In Figs.~\ref{n_odd} and \ref{n_even}, we have also shown the analytic 
results obtained in subsections A and B for the case of a fermion mass 
$M = 10$ TeV. For the case of odd $N$, shown in Fig.~\ref{n_odd}, we 
have plotted the analytic results obtained from Eqs.~\eqref{eq: de 
scaling overdamped} and \eqref{eq: de odd N}, represented by the green 
and orange lines respectively. We see that there is good agreement 
between the analytic formulae and our numerical results. For the 
lightest moduli, the $d_e(m_\phi )$ lines are in the overdamped regime 
and the slope closely follows the one predicted by Eq.~\eqref{eq: de 
scaling overdamped} (green line). As the mass increases, the line 
eventually approaches the region where all solutions fall into the 
central minimum. In this regime the $d_e(m_\phi)$ line tracks Eq. 
\eqref{eq: de odd N} (orange line), as expected.

For the plots in Fig. \ref{n_even} where $N$ is even, the behavior for 
the lowest masses $m_\phi$ again follows Eq.~\eqref{eq: de scaling 
overdamped}. According to Eq.~\eqref{eq: de scaling overdamped} and Eq. 
\eqref{eq: de scaling underdamped}, when $\eta \sim 1$ we should expect 
that the $d_e(m_\phi)$ line jumps while maintaining the same slope. This 
is exactly what occurs. For larger values of $m_\phi$ the lower 
$d_e(m_\phi)$ line merges with upper one.  This occurs because in the 
\new{($\eta > 1$)} region, as mentioned 
earlier, essentially all initial conditions lead to $\phi_{s} = {\pi 
f}/{N}$ as an initial condition.  Therefore the parameter space 
populated by the model is essentially independent of the initial 
misalignment angle.

In Figs.~\ref{n_odd} and~\ref{n_even}, we see the preferred parameter 
space of our model for various choices of $M$ and $N$.  From these 
figures, it is clear that our models span a wide range of $d_e$ and 
$m_\phi$.  Larger values of $N$ are currently being probed by ongoing 
experiments while smaller values of $N$ are within reach of future 
experiments.  Much as how the QCD axion line provides a goalpost for 
axion experiments, our small $N$ models constitute a well-motivated 
scenario that future experiments should aim to reach.

\section{Conclusions}
\label{Sec: conclusion}

Light moduli are one of the most attractive dark matter candidates. Not 
only are they ubiquitous in ultraviolet-complete models such as string 
theory, but the misalignment mechanism provides a simple explanation for 
their abundance. Despite these appealing features, detecting modulus 
dark matter remains a challenge. The problem is that the large coupling 
required to find them tends to be at odds with their very light mass, 
resulting in a hierarchy problem. In this article, we have shown 
that a nonlinearly realized $\mathbb{Z}_N$ symmetry can naturally 
protect the mass of a modulus against radiative corrections, provided 
the modulus itself transforms nonlinearly under the $\mathbb{Z}_N$.  It 
remains an intriguing open question whether a model that exhibits these 
features can be constructed within a string theory framework.

Much like the QCD axion, these models have a region of parameter space 
in which they naturally reproduce the observed dark matter abundance via 
the misalignment mechanism.  
The regions of parameter space populated by these models were shown in 
Figs.~\ref{n_odd} and \ref{n_even} for a few values of $N$. We see from 
these figures that future experiments searching for chronovariance of
the fine structure constant are expected to probe a large part of the 
preferred parameter space. This class of models is both simple and 
testable and provides additional theory motivation for exciting new 
quantum limited experiments.

\section*{Acknowledgements}

We thank Elizabeth Egbert and Junwu Huang for useful discussions. ZC and 
AD are supported in part by the National Science Foundation under Grant 
Number PHY-1914731. ZC is also supported in part by the US-Israeli BSF 
Grant 2018236. DB and AH are supported in part by the NSF under Grant 
No. PHY-1914480 and by the Maryland Center for Fundamental Physics 
(MCFP).

\appendix
\section{Finite Temperature Effects}
\label{Appendix: Thermal calculations}
 \new{In this appendix we obtain the contribution to the potential of 
the modulus from finite temperature effects. For temperatures $T\gtrsim 
M$, the leading contribution arises from the free energy contribution of 
the vector-like fermion $\Psi$ through the dependence of its mass on the 
value of the modulus, as given in Eq.~\eqref{eq. vector-like mass}. This 
contribution is represented by the one-loop diagram shown in 
Fig.~\ref{fig: fermion}.
 \begin{figure}[H] 
    \centering
    \includegraphics[width=0.3\linewidth]{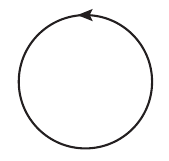}
    \caption{Diagram representing the leading contribution to the potential of the modulus for temperatures $T\gtrsim M$.}
    \label{fig: fermion}
 \end{figure}
 The form of this contribution is 
well-known~\cite{Laine_book,Kapusta_book} and is given by the integral
 \begin{equation}
    F_{f} =  \frac{2T^4}{\pi^2} \int\limits^{\infty}_{0}dx x^2 \ln{\left(1+\exp{-\sqrt{x^2+\left(\frac{M_{N}}{T}\right)^2}}\right)}
\end{equation}
 We can extract the effect on the potential of the modulus by Taylor 
expanding the potential around $M_{N} = M$ and keeping the correction of 
order $O(\epsilon)$,
 \begin{eqnarray}\label{eq: fermion contribution}
   V_T(\varphi,T) \supset  -\epsilon M \frac{d F_{f}}{dM} \cos{\frac{\varphi}{f}} .
\end{eqnarray}
 This leads to
 \begin{equation} \label{eq: fermion contribution explicit}
    -\epsilon M \frac{d F_{f}}{dM} \cos{\frac{\varphi}{f}} \approx  \left\{ \begin{matrix}
     \frac{\epsilon M^2 }{ 6} T^2 \cos{\frac{\varphi}{f}} && T > M\\
    \\
    \frac{4\epsilon M^{\frac{5}{2}} T^{\frac{3}{2}}}{(2\pi)^{\frac{3}{2}}} e^{-\frac{M}{T}} \cos{\frac{\varphi}{f}}&& T\lesssim M 
    \end{matrix} \right. \;.
\end{equation}
 Another type of contribution arises from 
the dependence of the free energy of the universe on the hypercharge 
gauge coupling, which in turn is a function of the VEV of the modulus,
 \begin{equation}
\label{phi-dependence} 
{g'}^2(\varphi) \supset \frac{\epsilon g'^{4} }{ \pi^2} \int\limits_{0}^{1} dx \frac{x(1-x)}{1+\left(\frac{T}{M}\right)^2 x(1-x)} \cos{\frac{\varphi}{f}} .
 \end{equation}
 From the above expression we can determine the dependence of this 
effect on the temperature,
 \begin{equation} \label{eq: gauge coupling correction}
    {g'}^2(\varphi) \supset  \left\{ \begin{matrix}
     \frac{\epsilon g'^{4} }{ \pi^2} \left(\frac{M}{T}\right)^2 \cos{\frac{\varphi}{f}} && T\gg M\\
    \\
    \frac{\epsilon g'^{4} }{6\pi^2}  \cos{\frac{\varphi}{f}} && T\ll M 
    \end{matrix} \right. \;.
 \end{equation}}
 The leading finite temperature contributions to the free energy of the 
universe that involve the hypercharge gauge coupling arise at two loops. 
The problem therefore reduces to computing these two loop diagrams. 
There are three types of diagrams labelled by $a$, $b$ and $c$ that 
contribute at this order,
 \new{
 \begin{equation} 
 \label{eq: phi_potential_2}
 \begin{aligned}
V_{T}(\varphi,T) &=  \frac{2\epsilon \alpha q^{2}_{F} u(\frac{T}{M})}{3 \pi \cos^{2}\theta_{W} } \left( \sum_{i} F_{a}^{i} + F_{b} + F_{c} \right) \cos \left( \frac{\varphi}{f} \right) \; ,
 \end{aligned}
 \end{equation}
  where
 \begin{equation}
     u(y) = \int\limits_{0}^{1} dx \frac{6 x(1-x)}{1+y^2 x(1-x)} \; .
 \end{equation}}
 
\begin{figure}[H] 
    \centering
    \includegraphics[width=0.9\linewidth]{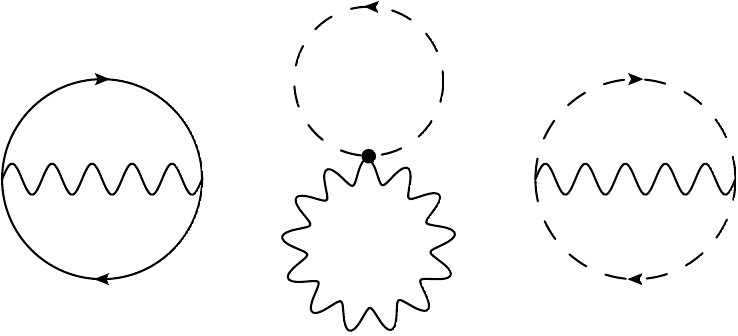}
    \caption{Diagrams contributing to the effective potential. The first diagram on the left corresponds to the term $F_{a}^{i}$, the next to $F_{b}$ and the last to $F_{c}$ as defined in Eq.~(\ref{eq: phi_potential_2}).}
    \label{fig: diagrams}
 \end{figure}

 The diagrams that give rise to $F_{a}^{i}$, $F_{b}$ and $F_{c}$ are 
shown in Fig. \ref{fig: diagrams}. Here the index $i$ runs over all 
quark and lepton flavors.
 Using the methods of thermal field 
theory~\cite{Laine_book,Kapusta_book} the first diagram, which 
represents the correction to the free energy from diagrams involving the 
SM fermions, can be evaluated 
 as
 \begin{equation}
\begin{aligned} \label{eq: fai}
    F_{a}^{i} = -\frac{g'^{2}q^{2}_{i}}{2} \sumint\limits_{\{P,R\}Q} &\frac{Tr\{\tilde{\gamma}_{\mu} (-i\slashed{P}+m_{i})\tilde{\gamma}_{\mu}(-i \slashed{R}+m_{i}) \}}{Q^{2} (P^{2}+m^{2}_{i})(R^{2}+m^{2}_{i})} \times \\
    &\times \delta^{(4)}(P+Q-R) 
\end{aligned}    
 \end{equation}
 where we follow the conventions of \cite{Laine_book}. In these 
conventions the $\tilde{\gamma}_{\mu}$ represent the Euclidean Dirac 
matrices,
 \begin{equation}
    \tilde{\gamma}_{0} \equiv \gamma^{0} \, , \qquad \tilde{\gamma}_{k} \equiv -i \gamma^{k} \, , \qquad k=1, ...,d \, .
 \end{equation}
 The momenta in the loop integrals are in Euclidean space and have 
components $P = (\omega^{i}_{n},\Vec{p})$, where $i = b,f$. The symbol 
$\sumint$ refers to integration over the spatial momenta along with 
summation over the Matsubara frequencies, which are defined as
 \begin{equation}
    \omega^{f}_{n} = 2 \pi T (n+\frac{1}{2}) \, , \qquad \omega^{b}_{n} = 2 \pi T n
 \end{equation}
 for fermions and bosons respectively.  If the momenta below the symbol $\sumint$ appear 
between curly brackets $\{...\}$ it corresponds to summation and 
integration over fermionic boundary conditions while the absence of the brackets 
indicate bosonic boundary conditions. After the usual Dirac algebra we 
can express Eq.~(\ref{eq: fai}) in terms of known standard integrals 
\cite{Arnold_pureQCD,Arnold_QCD},
 \begin{equation} \label{eq: Fa}
\begin{aligned}
    F_{a}^{i} = &-\frac{d}{2} (\frac{d}{2} -1) g'^{2} q^{2}_{i}  \left\{ 2 \sumint\limits_{\{P\}} \frac{1}{P^{2}+m^{2}_{i}} \sumint\limits_{Q} \frac{1}{Q^{2}} -\left[ \sumint\limits_{\{P\}} \frac{1}{P^{2}+m^{2}_{i}} \right]^{2} \right. \\
    &+\frac{4m^{2}_{i}}{d-2} \left.  \sumint\limits_{\{P,R\}} \frac{1}{(P-R)^{2}(P^{2}+m^{2}_{i})(R^{2}+m^{2}_{i})}  \right\} \, ,
\end{aligned}    
 \end{equation}
 where $d = 4 - 2\epsilon$. In the $T/m \to \infty$ limit the last 
integral vanishes~\cite{Arnold_QCD}. The remaining integrals are well 
known,
 \begin{equation}
    \sumint\limits_{Q} \frac{1}{Q^{2}} = \frac{T^{2}}{12} \qquad \text{and} \qquad \sumint\limits_{\{P\}} \frac{1}{P^{2}} = -\frac{T^{2}}{24}.  
 \end{equation}
 Putting these together we arrive at the result in the high 
temperature/low mass limit,
 \begin{equation} \label{eq: Fa low mass}
   F_{a}^{i} \approx   \frac{5 \pi \alpha  q^{2}_{i}}{72 \cos^{2}\theta_{W}} T^{4} \; ,
 \end{equation}
 where we have expressed $g'$ in terms of the fine structure constant 
$\alpha$ and the Weinberg angle $\theta_{W}$. Since the $q_{i}$ 
represent the hypercharges of the SM fermions and the heavy fermion $\Psi$ from our sector, after including all the 
contributing particles we obtain a factor of $\sum_{i} q^{2}_{i} = 6$.

The second and the third diagrams represent contributions to the free 
energy from the Higgs doublet. We focus on the high temperature/low mass 
limit prior to electroweak symmetry breaking. The contribution to the 
amplitude from the second diagram can be written as the product of the 
amplitude of two independent one loop diagrams,
 \begin{equation}
    F_{b} = \frac{g'^{2}}{2}g^{\mu}_{\mu}  \sumint\limits_{P} \frac{1}{P^{2}+m^{2}} \sumint\limits_{Q} \frac{1}{Q^{2}} \; . 
 \end{equation}
 In the $T/m \to \infty$ limit we obtain,
 \begin{equation}
    F_{b} \approx  \frac{\pi \alpha }{18\cos^{2}\theta_{W}} T^{4} \; .
 \end{equation}
 Finally, the third diagram has a form very similar to the first 
\eqref{eq: Fa}, except that now all the integrals are bosonic,
 \begin{equation}
\begin{aligned}
    F_{c} &= \frac{g'^2}{4}  \sumint\limits_{P,Q,R} \frac{(P+R)^{2}}{(P^{2}+m^{2})(R^{2}+m^{2})Q^{2}} \delta^{4}(R-P-Q)\\
    &\approx \frac{\pi \alpha}{48 \cos^{2}\theta_{W}} T^{4} \; .
\end{aligned}
 \end{equation}
 \new{ Now, using Eqs. \eqref{eq: fermion contribution explicit} and 
\eqref{eq: phi_potential_2} we put all the pieces together and see that 
for $T \gg M$, the temperature dependent contribution simplifies to,
 \begin{equation}
 V_{T}(\varphi,T) =  \left(\frac{71\epsilon \alpha^{2} }{36 
\cos^{4}\theta_{W}} + \frac{\epsilon}{6} \right) T^{2}M^{2}\cos \left( \frac{\varphi}{f} \right) .
 \end{equation}
 When the temperature falls below $T\sim M$, the contribution given by 
Eq.~\eqref{eq: fermion contribution explicit} becomes exponentially 
suppressed while the correction to the gauge coupling from 
Eq.~\eqref{eq: gauge coupling correction} becomes independent of 
temperature. Additionally, the vector-like fermion $\Psi$ of our sector no longer contributes to the Eq. \eqref{eq: Fa low mass}, resulting in $\sum_{i} q^{2}_{i} = 5$. Therefore, when the temperature enters the range $M 
\gtrsim T \gtrsim 100$ GeV, the potential takes the form
 \begin{equation} 
 \label{eq: phi_potential_2 high t low m}
\begin{aligned}
V_{T}(\varphi,T) &= \left(\frac{61\epsilon \alpha^{2} q^{2}_{F}  }{216 \cos^{4}\theta_{W}} T^{4}+\frac{4\epsilon M^{\frac{5}{2}} T^{\frac{3}{2}}}{(2\pi)^{\frac{3}{2}}} e^{-\frac{M}{T}}\right)\cos \left( \frac{\varphi}{f} \right).
\end{aligned}
 \end{equation}}
 After electroweak symmetry breaking, we can neglect the contribution 
from the $Z$ boson as it is suppressed almost immediately. To get the 
contribution from photons we multiply the temperature dependent part in 
Eq.~\eqref{eq: phi_potential_2} by $\cos^{4} \theta_{W}$, which effectively 
replaces the hypercharge gauge coupling $g'$ by $e$
 \begin{equation} \label{eq: phi_potential_2 after SSB}
\begin{aligned}
V_{T}(\varphi,T) &= \epsilon \alpha^{2} q^{2}_{F} q^{2}_{eff}(T)  T^{4}\cos \left( \frac{\varphi}{f} \right).
\end{aligned}
 \end{equation}
 Here $q^{2}_{eff}(T)$ is a coefficient that changes with the number of 
species that contribute to the temperature dependent potential. As the 
universe cools down the number of species that contribute to the 
potential drops as heavier fields go out of the bath. This in turn, 
increases the relative strength of Hubble friction. However, this 
effect is partially compensated for by the rise in the temperature when 
heavy fields exit the thermal bath.

Therefore, the thermal potential does not change qualitatively until the 
temperature drops below the mass of the lightest charged particle, the 
electron, so that $T \lesssim m_{e}$. However, in order for this effect 
to play any role in the evolution of the modulus $\varphi$, the thermal 
part $V_T(\varphi,T)$ has to be significant at this point in time, which 
requires \begin{equation}
    \eta \gtrsim \eta_0,
\end{equation}
where $\eta_0$ was defined in Eq.~\eqref{eq: eta0 value}. On the top of that, $V_T(\varphi,T)$ has to dominate over $V_0(\varphi)$ at $T\sim m_{e}$. Hence, we have to satisfy

\new{\begin{equation} \label{eq: conditions LT full}
    T_{s} \lesssim 1 \text{MeV} \qquad \text{and} \qquad \eta \gtrsim \eta_{0}
\end{equation}
}
 If we additionally require that these conditions overlap with the 
parameter space not excluded by experiments, we conclude that for $M 
\gtrsim 1\text{ TeV}$ no solutions exist that satisfy the above 
criteria simultaneously. It therefore suffices to consider the high 
temperature limit of the temperature dependent potential.

\section{Form of the Potential at the Onset of Oscillations}
\label{Appendix: amplitude underdamped}

In section \ref{Sec: history} we argued that in the case when $N$ is 
even and we are in the underdamped scenario, \new{$\eta > 1$}, the energy 
density is well approximated by Eq.~\eqref{Eq: underdamped rho}. This 
formula assumes that field only begins the final stage of oscillations 
once the finite temperature contributions to the potential are small, so 
that that the potential is well approximated by its zero temperature 
form. In order to verify that this condition holds, we will show that 
for \new{most of} the range of parameters considered in Fig.~\ref{n_even}, the height 
difference between the the central maximum and the closest minimum is 
already within $10\%$ of its final value at the time that the 
oscillations begin.
Recall that the potential is proportional to the sum of two cosines
 \begin{equation} \label{eq: f definition}
\begin{aligned}
    V(\varphi,T) \propto &\, \frac{N^{2} m^2(T)}{m_{\phi}^{2}}\cos \left(\frac{ \varphi}{f}\right)+ \cos \left(\frac{N \varphi}{f}\right) \\
    &= \frac{N^{2} T^4}{T_{tr}^{4}}\cos \left(\frac{ \varphi}{f}\right)+ \cos \left(\frac{N \varphi}{f}\right)
\end{aligned}
 \end{equation}
\new{where $m^2(T)$ is defined in Eq.~\eqref{eq: m(T) definition} and $T_{tr}$ is
the temperature at which $V_0(\varphi)$ and $V_T(\varphi,T)$ have 
quadratic terms of the same magnitude at $\varphi = \pi f$,
 \begin{equation} \label{eq: Teq1}
    T_{tr} = \left(\frac{216 \cos^{4} \theta_{W} }{61 \alpha^{2}} \right)^{1/4} \frac{\sqrt{m_{\phi} f}}{\epsilon^{1/4}} = \sqrt{6} T_{s}\eta^{-1/4}.
 \end{equation}}
 Naively, oscillations 
will begin at a temperature $\tilde{T}$ when 
 \begin{equation} 
 \label{eq: Ttilde def}
   \tilde{m}^{2} \equiv  m_{\phi}^2 - m^2(\tilde{T}) = 9 H^2
 \end{equation}
  This equation defines $ \tilde{m}$, the effective mass at temperature 
$\tilde{T}$. Eq.~\eqref{eq: Ttilde def} may be rewritten as,
 \begin{equation} 
 \label{eq: Ttr and Ttilde}
   \frac{T_{tr}^{4}}{ \tilde{T}^4} =  \frac{m_{\phi}^2}{m^2(\tilde{T})} =1+ 9\frac{ H^2}{m^2(\tilde{T})} = 1 + \frac{36}{\eta},
 \end{equation}
 At temperature $\tilde{T}$, Hubble friction is sufficiently small for 
the field $\phi$ to roll down. However, due to the small gradient near the top 
of the potential, the moment it reaches the bottom of the potential is 
delayed by approximately~\cite{Lyth_1992}
 \begin{equation} \label{eq: delta t}
    \Delta t = \frac{2}{3 m_{\phi}} \log \frac{1}{x} = t_{s}\log \frac{1}{x}, 
 \end{equation}
 where $x = 1- \frac{N \phi }{\pi f}$ is a measure of how close to the 
maximum the field is when oscillations about the zero-temperature 
minimum begin and $t_{s}$ is the time at which oscillations would start 
in the overdamped scenario. The value of $x$ can be obtained from Eq. 
\eqref{eq: phi initial value underdamped} as
 \new{
 \begin{equation}
x = \frac{N}{\sqrt3}\sqrt{\frac{m(T_{osc})}{m_{\phi}}} \left( \frac{T_s}{T_{o}} \right)^{\frac{3}{2}}  \;.    \end{equation}
 }
 Before moving on, it is important to point out that the Eq.~\eqref{eq: 
delta t} provides a good qualitative explanation of the $\phi$ 
evolution, but does not lead to Eq.~\eqref{Eq: underdamped rho}, which 
is purely empirical. Now, with Eq.~\eqref{eq: delta t} we see that the 
time at which oscillations effectively start is
 \begin{equation}
   t_{eff} = \tilde{t}+t_{s}\log \frac{1}{x} ,
 \end{equation}
 where $\tilde{t}$ is a time when $T=\tilde{T}$. Using the fact that 
this transition always occurs during the radiation dominated epoch, when 
$T\propto t^{-1/2}$, we can write
 \begin{equation}
   \frac{T_{s}^{2}}{T_{eff}^{2}} = \frac{T_{s}^{2}}{\tilde{T}^{2}}+\log \frac{1}{x} .
 \end{equation}
 Then, with help from Eq.~\eqref{eq: Teq1} and Eq.~\eqref{eq: Ttr and 
Ttilde}, we can write
 \begin{equation} 
 \label{eq: T_tr/T_eff}
   \frac{T_{eff}^{4}}{T_{tr}^{4}}= \frac{\eta}{36} \left( \sqrt{1+\frac{\eta}{36}}+\log \frac{1}{x} \right)^{-2} .
 \end{equation}
 The ratio ${T_{eff}^{4}}/{T_{tr}^{4}}$ determines the shape of the 
potential in Eq.~\eqref{eq: f definition} at $T=T_{eff}$. Eq.~\eqref{eq: 
T_tr/T_eff} relates this ratio to the parameter $\eta$. Using these 
equations we can determine the ratio of the height difference between 
the central maximum and the closest minimum at $T=T_{eff}$ relative to 
the same height difference at zero temperature,
 \begin{equation}
\begin{aligned}
   \frac{\Delta h}{h} &=\frac{V(\pi f,T_{eff})-V(\pi f + \pi f/N,T_{eff}) }{V(\pi f ,0)-V(\pi f + \pi f/N,0)} = \\ 
   &= 1-(1-\cos{\frac{\pi}{N}})\frac{N^{2}T_{eff}^{4}}{2 T_{tr}^{4}} = \\
    &= 1-(1-\cos{\frac{\pi}{N}})\frac{N^2 \eta}{72} \left( \sqrt{1+\frac{\eta}{36}}+\log \frac{1}{x} \right)^{-2} \;.
\end{aligned}    
 \end{equation} 
 \new{Since the lowest value of $\log \frac{1}{x}$ for both $N=6$ and 
$N=10$ is $\log \frac{1}{x} \approx 5$, in order to satisfy $\Delta h/h 
> 0.9$ we need $\eta\ll 70$. This condition is easily fulfilled for $N=6$ 
as $\eta \lesssim 40 $. However, for $N=10$ it is satisfied only for 
smaller values of the mass of the modulus, $m_{\phi} \lesssim 10^{-12}$ 
eV. Nevertheless, as evident from the numerical simulations presented in 
Fig. \ref{n_even}, Eq.~(\ref{eq: de scaling underdamped}) continues to 
give a good approximation to the actual result even when this condition 
is not satisfied.
}

\section{Parametric Resonance}

One may worry that decays of $\phi$ into photons may deplete the 
abundance of dark matter, particularly if the decay rate is enhanced by 
parametric resonance. In this appendix we consider this question. We 
find that in the region of parameter space of interest, the condition 
for parametric resonance is not satisfied. Therefore these decays are 
slow on cosmological time scales and the abundance of dark matter is not 
affected.

The Lagrangian for the coupling of $\phi$ to photons can be written as
 \begin{equation}
   \mathcal{L}=-\frac{1}{4 g^2(\phi)}F^2~~~\text{where}~~~~ \frac{1}{g^2(\phi)} = \frac{1}{e^2} + \frac{\epsilon}{6 \pi^2}\frac{\phi}{f}.
 \end{equation}
 The resulting equation of motion for the photon is given by 
 \begin{equation}
   g^2(\phi) \partial_\mu\{g^{-2}(\phi)\} F^{\mu\nu}+\Box A^\nu- \partial^\nu \partial_\mu A^\mu=0 \;.
 \end{equation}
 Working in the gauge where $A_0,\partial_\mu A^\mu =0$ this becomes,
 \begin{equation}
  \Box A^\nu = -e^2 \frac{\epsilon}{6 \pi^2} \frac{\dot{\phi}}{f}\partial_0 A^\nu  .
 \end{equation}
 This leads to the modified dispersion relation,
 \begin{equation}
    (\omega^2 - k^2)= \frac{\epsilon e^2}{6 \pi^2}  \frac{\dot{\phi} \omega}{f}.
 \end{equation}

 Because the plasma mass of the photon is large at the time of the 
phase transition, we are interested in parametric resonance at times 
when $\phi/f \ll 1$. Perturbatively expanding the time-dependent 
dispersion relation, we obtain
 \begin{equation}
    k=\omega-\delta k ~~~~~~~~~~~~~~~~~\text{where}~~~~~~~~~~ \delta k = \frac{\epsilon e^2}{12 \pi^2} \frac{\dot{\phi}}{f}.
 \end{equation}
 When the $\phi$ condensate decays, the resulting photons have the 
frequency $\omega=m_\phi/2$. In phase space, the photons are emitted 
into a thin shell centered around $m_\phi/2$ with width $2\delta k$. The 
occupation number of photons $\tilde{n}_\gamma^k$ with momentum $k$ can 
be related to the number density of photons $n_\gamma$ as
 \begin{equation}
\tilde{n}_\gamma ^k = \frac{n_\gamma}{4 \pi k^2 2\delta k}.
 \end{equation}
 The integrated Boltzmann equation for the production of photons from 
the decay of the $\phi$ condensate in the Bose enhanced regime 
$(\tilde{n}_\gamma ^k \gg 1)$ is given by,
 \begin{equation}
 \label{BE}
\frac{d}{dt}(a^3 n_\gamma)=2 a^3 \Gamma (1+ 2 \tilde{n}_\gamma ^k) n_\phi \approx  4 a^3  \Gamma  \tilde{n}_\gamma ^k n_\phi = a^3\frac{\Gamma n_\phi}{2 \pi k^2 \delta k} n_\gamma
 \end{equation}
 where $\Gamma$ denotes the decay rate of the condensate into photons at zero 
background photon density. This decay rate $\Gamma$ can be estimated as
 \begin{eqnarray}
    \Gamma\sim \left(\frac{e^2 \epsilon}{6 \pi^2}\right)^2\frac{ m_\phi^3}{f^2}. \\
    \qquad \nonumber
 \end{eqnarray}
 For all of the benchmark points in Table~\ref{Tab: values}, the 
lifetime is greater than $10^{75}$ years. This is much greater than the 
age of the universe. Therefore, in the absence of parametric resonance, 
decays of $\phi$ can safely be neglected.

 The parametrically-resonant decay rate can be easily read off from Eq. 
(\ref{BE}) as
 \begin{equation}
    \Gamma_{res}= \frac{\Gamma n_\phi}{2 \pi k^2 \delta k} = \frac{2 e^2 \epsilon m_\phi}{3 \pi^3 } \frac{\phi}{f},
 \end{equation}
 where we have used $n_\phi = m_\phi \phi ^2$ and $\dot{\phi}/\phi = 
m_\phi$. In an expanding universe the proper momentum is redshifting as 
$k=k_{comoving}/a$. This means that the radius of the thin momentum 
shell is shrinking at the rate of $k H$. Since the momentum shell has a 
finite width $2\delta k$, it takes time $t_k=2\delta k /(k H)$ for an 
emitted photon to redshift outside the momentum shell. In order to avoid 
Bose enhancement of decays in the expanding universe, this time scale 
$t_k$ has to be parametrically smaller than the decay rate 
$\Gamma_{res}$ at the time of decay,
 \begin{equation}\label{condition}
\frac{2\delta k}{k H}  \ll \frac{1}{\Gamma_{res}}.
 \end{equation}
 At early times prior to photon decoupling, the photon has a plasma mass 
which is larger than $m_\phi$. Consequently, decays of $\phi$ into 
photons are kinematically forbidden until after decoupling. This means 
that the condition in Eq.~(\ref{condition}) only has to be satisfied at 
times after $T_{decoupling}\approx 0.3$ eV to avoid parametric 
resonance. Since decoupling happens during matter domination, the Hubble 
expansion at these times can be related to the field $\phi$ as
 \begin{equation}
H=\sqrt{\rho_m}/(3 M_{pl})=m_\phi\phi/(3 M_{pl}\sqrt{r}),
 \end{equation}
 where $r$ is the fraction of dark matter constituted by $\phi$.
Then, the condition to avoid parametric resonance becomes, 
 \begin{equation}
\frac{2 \sqrt{r} \epsilon^2 e^4}{3 \pi^5} \frac{M_{pl}\phi}{f^2} \ll 1 . 
 \end{equation} 
 This condition is easily satisfied because the amplitude of the field 
$\phi$ has already undergone significant damping by the time of 
recombination.

\bibliography{bibliography}{}

\begin{thebibliography}{10}

\bibitem{Aad:2012tfa}
Georges Aad et~al.
\newblock {Observation of a new particle in the search for the Standard Model
  Higgs boson with the ATLAS detector at the LHC}.
\newblock {\em Phys. Lett. B}, 716:1--29, 2012.

\bibitem{Chatrchyan:2012ufa}
Serguei Chatrchyan et~al.
\newblock {Observation of a New Boson at a Mass of 125 GeV with the CMS
  Experiment at the LHC}.
\newblock {\em Phys. Lett. B}, 716:30--61, 2012.

\bibitem{Aghanim:2018eyx}
N.~Aghanim et~al.
\newblock {Planck 2018 results. VI. Cosmological parameters}.
\newblock 7 2018.

\bibitem{Goldberger:1999uk}
Walter~D. Goldberger and Mark~B. Wise.
\newblock {Modulus stabilization with bulk fields}.
\newblock {\em Phys. Rev. Lett.}, 83:4922--4925, 1999.

\bibitem{Arvanitaki:2009fg}
Asimina Arvanitaki, Savas Dimopoulos, Sergei Dubovsky, Nemanja Kaloper, and
  John March-Russell.
\newblock {String Axiverse}.
\newblock {\em Phys. Rev. D}, 81:123530, 2010.

\bibitem{Chacko:2012sy}
Zackaria Chacko and Rashmish~K. Mishra.
\newblock {Effective Theory of a Light Dilaton}.
\newblock {\em Phys. Rev. D}, 87(11):115006, 2013.

\bibitem{Coradeschi:2013gda}
Francesco Coradeschi, Paolo Lodone, Duccio Pappadopulo, Riccardo Rattazzi, and
  Lorenzo Vitale.
\newblock {A naturally light dilaton}.
\newblock {\em JHEP}, 11:057, 2013.

\bibitem{Kane:2015jia}
Gordon Kane, Kuver Sinha, and Scott Watson.
\newblock {Cosmological Moduli and the Post-Inflationary Universe: A Critical
  Review}.
\newblock {\em Int. J. Mod. Phys. D}, 24(08):1530022, 2015.

\bibitem{Linde_1987}
Andrei~D. Linde.
\newblock {Inflation and Axion Cosmology}.
\newblock {\em Phys. Lett.}, B201:437--439, 1988.

\bibitem{Damour2010_convention}
Thibault Damour and John~F. Donoghue.
\newblock Equivalence principle violations and couplings of a light dilaton.
\newblock {\em Phys. Rev. D}, 82:084033, Oct 2010.

\bibitem{Damour:2010rm}
Thibault Damour and John~F. Donoghue.
\newblock {Phenomenology of the Equivalence Principle with Light Scalars}.
\newblock {\em Class. Quant. Grav.}, 27:202001, 2010.

\bibitem{Damour:2012rc}
Thibault Damour.
\newblock {Theoretical Aspects of the Equivalence Principle}.
\newblock {\em Class. Quant. Grav.}, 29:184001, 2012.

\bibitem{Khmelnitsky:2013lxt}
Andrei Khmelnitsky and Valery Rubakov.
\newblock {Pulsar timing signal from ultralight scalar dark matter}.
\newblock {\em JCAP}, 02:019, 2014.

\bibitem{Stadnik:2014tta}
Y.V. Stadnik and V.V. Flambaum.
\newblock {Searching for dark matter and variation of fundamental constants
  with laser and maser interferometry}.
\newblock {\em Phys. Rev. Lett.}, 114:161301, 2015.

\bibitem{Arvanitaki:2015iga}
Asimina Arvanitaki, Savas Dimopoulos, and Ken Van~Tilburg.
\newblock {Sound of Dark Matter: Searching for Light Scalars with Resonant-Mass
  Detectors}.
\newblock {\em Phys. Rev. Lett.}, 116(3):031102, 2016.

\bibitem{Graham:2015ouw}
Peter~W. Graham, Igor~G. Irastorza, Steven~K. Lamoreaux, Axel Lindner, and
  Karl~A. van Bibber.
\newblock {Experimental Searches for the Axion and Axion-Like Particles}.
\newblock {\em Ann. Rev. Nucl. Part. Sci.}, 65:485--514, 2015.

\bibitem{Graham:2015ifn}
Peter~W. Graham, David~E. Kaplan, Jeremy Mardon, Surjeet Rajendran, and
  William~A. Terrano.
\newblock {Dark Matter Direct Detection with Accelerometers}.
\newblock {\em Phys. Rev. D}, 93(7):075029, 2016.

\bibitem{Stadnik:2015xbn}
Y.V. Stadnik and V.V. Flambaum.
\newblock {Enhanced effects of variation of the fundamental constants in laser
  interferometers and application to dark matter detection}.
\newblock {\em Phys. Rev. A}, 93(6):063630, 2016.

\bibitem{Delaunay:2016brc}
C\'edric Delaunay, Roee Ozeri, Gilad Perez, and Yotam Soreq.
\newblock {Probing Atomic Higgs-like Forces at the Precision Frontier}.
\newblock {\em Phys. Rev. D}, 96(9):093001, 2017.

\bibitem{Berlin:2016woy}
Asher Berlin.
\newblock {Neutrino Oscillations as a Probe of Light Scalar Dark Matter}.
\newblock {\em Phys. Rev. Lett.}, 117(23):231801, 2016.

\bibitem{Geraci:2016fva}
Andrew~A. Geraci and Andrei Derevianko.
\newblock {Sensitivity of atom interferometry to ultralight scalar field dark
  matter}.
\newblock {\em Phys. Rev. Lett.}, 117(26):261301, 2016.

\bibitem{Krnjaic:2017zlz}
Gordan Krnjaic, Pedro~A.N. Machado, and Lina Necib.
\newblock {Distorted neutrino oscillations from time varying cosmic fields}.
\newblock {\em Phys. Rev. D}, 97(7):075017, 2018.

\bibitem{Roberts:2017hla}
Benjamin~M. Roberts, Geoffrey Blewitt, Conner Dailey, Mac Murphy, Maxim
  Pospelov, Alex Rollings, Jeff Sherman, Wyatt Williams, and Andrei Derevianko.
\newblock {Search for domain wall dark matter with atomic clocks on board
  global positioning system satellites}.
\newblock {\em Nature Commun.}, 8(1):1195, 2017.

\bibitem{Arvanitaki:2017nhi}
Asimina Arvanitaki, Savas Dimopoulos, and Ken Van~Tilburg.
\newblock {Resonant absorption of bosonic dark matter in molecules}.
\newblock {\em Phys. Rev. X}, 8(4):041001, 2018.

\bibitem{DeRocco:2018jwe}
William DeRocco and Anson Hook.
\newblock {Axion interferometry}.
\newblock {\em Phys. Rev. D}, 98(3):035021, 2018.

\bibitem{Geraci:2018fax}
Andrew~A. Geraci, Colin Bradley, Dongfeng Gao, Jonathan Weinstein, and Andrei
  Derevianko.
\newblock {Searching for Ultralight Dark Matter with Optical Cavities}.
\newblock {\em Phys. Rev. Lett.}, 123(3):031304, 2019.

\bibitem{Irastorza:2018dyq}
Igor~G. Irastorza and Javier Redondo.
\newblock {New experimental approaches in the search for axion-like particles}.
\newblock {\em Prog. Part. Nucl. Phys.}, 102:89--159, 2018.

\bibitem{Carney:2019cio}
Daniel Carney, Anson Hook, Zhen Liu, Jacob~M. Taylor, and Yue Zhao.
\newblock {Ultralight Dark Matter Detection with Mechanical Quantum Sensors}.
\newblock 8 2019.

\bibitem{Guo:2019ker}
Huai-Ke Guo, Keith Riles, Feng-Wei Yang, and Yue Zhao.
\newblock {Searching for Dark Photon Dark Matter in LIGO O1 Data}.
\newblock {\em Commun. Phys.}, 2:155, 2019.

\bibitem{Grote:2019uvn}
H.~Grote and Y.V. Stadnik.
\newblock {Novel signatures of dark matter in laser-interferometric
  gravitational-wave detectors}.
\newblock {\em Phys. Rev. Res.}, 1(3):033187, 2019.

\bibitem{Dev:2020kgz}
Abhish Dev, Pedro~A.N. Machado, and Pablo Mart\'\i{}nez-Mirav\'e.
\newblock {Signatures of Ultralight Dark Matter in Neutrino Oscillation
  Experiments}.
\newblock 7 2020.

\bibitem{Stadnik:2020bfk}
Yevgeny~V. Stadnik.
\newblock {New bounds on macroscopic scalar-field topological defects from
  non-transient signatures due to environmental dependence and spatial
  variations of the fundamental constants}.
\newblock {\em Phys. Rev. D}, 102:115016, 2020.

\bibitem{Dirac:1937ti}
Paul A.~M. Dirac.
\newblock {The Cosmological constants}.
\newblock {\em Nature}, 139:323, 1937.

\bibitem{Chodos:1979vk}
Alan Chodos and Steven~L. Detweiler.
\newblock {Where Has the Fifth-Dimension Gone?}
\newblock {\em Phys. Rev. D}, 21:2167, 1980.

\bibitem{Terazawa:1981ga}
Hidezumi Terazawa.
\newblock {Cosmological Origin of Mass Scales}.
\newblock {\em Phys. Lett. B}, 101:43--47, 1981.

\bibitem{Bekenstein:1982eu}
J.~D. Bekenstein.
\newblock {Fine Structure Constant: Is It Really a Constant?}
\newblock {\em Phys. Rev. D}, 25:1527--1539, 1982.

\bibitem{Marciano:1983wy}
William~J. Marciano.
\newblock {Time Variation of the Fundamental 'Constants' and Kaluza-Klein
  Theories}.
\newblock {\em Phys. Rev. Lett.}, 52:489, 1984.

\bibitem{Anson1802}
Anson Hook.
\newblock Solving the hierarchy problem discretely.
\newblock {\em Phys. Rev. Lett.}, 120:261802, Jun 2018.

\bibitem{Arvanitaki2015}
Asimina Arvanitaki, Junwu Huang, and Ken Van~Tilburg.
\newblock Searching for dilaton dark matter with atomic clocks.
\newblock {\em Phys. Rev. D}, 91:015015, Jan 2015.

\bibitem{Arvanitaki2018}
Asimina Arvanitaki, Peter~W. Graham, Jason~M. Hogan, Surjeet Rajendran, and Ken
  Van~Tilburg.
\newblock Search for light scalar dark matter with atomic gravitational wave
  detectors.
\newblock {\em Phys. Rev. D}, 97:075020, Apr 2018.

\bibitem{VanTilburg2015}
Ken Van~Tilburg, Nathan Leefer, Lykourgos Bougas, and Dmitry Budker.
\newblock Search for ultralight scalar dark matter with atomic spectroscopy.
\newblock {\em Phys. Rev. Lett.}, 115:011802, Jun 2015.

\bibitem{Hess2016}
A.~Hees, J.~Gu\'ena, M.~Abgrall, S.~Bize, and P.~Wolf.
\newblock Searching for an oscillating massive scalar field as a dark matter
  candidate using atomic hyperfine frequency comparisons.
\newblock {\em Phys. Rev. Lett.}, 117:061301, Aug 2016.

\bibitem{Berge2017_MICROSCOPE}
Joel Bergé, Philippe Brax, Gilles Métris, Martin Pernot-Borràs, Pierre
  Touboul, and Jean-Philippe Uzan.
\newblock {MICROSCOPE Mission: First Constraints on the Violation of the Weak
  Equivalence Principle by a Light Scalar Dilaton}.
\newblock {\em Phys. Rev. Lett.}, 120(14):141101, 2018.

\bibitem{Hoyle1999_EP_CuPb}
G.~L. Smith, C.~D. Hoyle, J.~H. Gundlach, E.~G. Adelberger, B.~R. Heckel, and
  H.~E. Swanson.
\newblock Short-range tests of the equivalence principle.
\newblock {\em Phys. Rev. D}, 61:022001, Dec 1999.

\bibitem{Schlamminger2008_EP_BeTi}
S.~Schlamminger, K.-Y. Choi, T.~A. Wagner, J.~H. Gundlach, and E.~G.
  Adelberger.
\newblock Test of the equivalence principle using a rotating torsion balance.
\newblock {\em Phys. Rev. Lett.}, 100:041101, Jan 2008.

\bibitem{Baggio2005_AURIGA}
L.~Baggio, M.~Bignotto, M.~Bonaldi, M.~Cerdonio, L.~Conti, P.~Falferi,
  N.~Liguori, A.~Marin, R.~Mezzena, A.~Ortolan, S.~Poggi, G.~A. Prodi,
  F.~Salemi, G.~Soranzo, L.~Taffarello, G.~Vedovato, A.~Vinante, S.~Vitale, and
  J.~P. Zendri.
\newblock 3-mode detection for widening the bandwidth of resonant gravitational
  wave detectors.
\newblock {\em Phys. Rev. Lett.}, 94:241101, Jun 2005.

\bibitem{Kennedy:2020bac}
Colin~J. Kennedy, Eric Oelker, John~M. Robinson, Tobias Bothwell, Dhruv Kedar,
  William~R. Milner, G.~Edward Marti, Andrei Derevianko, and Jun Ye.
\newblock {Precision Metrology Meets Cosmology: Improved Constraints on
  Ultralight Dark Matter from Atom-Cavity Frequency Comparisons}.
\newblock {\em Phys. Rev. Lett.}, 125(20):201302, 2020.

\bibitem{Vermeulen:2021epa}
Sander~M. Vermeulen et~al.
\newblock {Direct limits for scalar field dark matter from a gravitational-wave
  detector}.
\newblock 3 2021.

\bibitem{nuclearclock}
P~G Thirolf, B~Seiferle, and L~von~der Wense.
\newblock The 229-thorium isomer: doorway to the road from the atomic clock to
  the nuclear clock.
\newblock {\em Journal of Physics B: Atomic, Molecular and Optical Physics},
  52(20):203001, sep 2019.

\bibitem{Coleman_MAGIS100}
Jon Coleman.
\newblock {Matter-wave Atomic Gradiometer InterferometricSensor (MAGIS-100) at
  Fermilab}.
\newblock {\em PoS}, ICHEP2018:021, 2019.

\bibitem{Badurina2019_AION}
L.~Badurina et~al.
\newblock {AION: An Atom Interferometer Observatory and Network}.
\newblock 2019.

\bibitem{Bertoldi2019_AEDGE}
Yousef~Abou El-Neaj et~al.
\newblock {AEDGE: Atomic Experiment for Dark Matter and Gravity Exploration in
  Space}.
\newblock 2019.

\bibitem{Stadnik:2015kia}
Y.V. Stadnik and V.V. Flambaum.
\newblock {Can dark matter induce cosmological evolution of the fundamental
  constants of Nature?}
\newblock {\em Phys. Rev. Lett.}, 115(20):201301, 2015.

\bibitem{Sibiryakov:2020eir}
Sergey Sibiryakov, Philip Sorensen, and Tien-Tien Yu.
\newblock {BBN constraints on universally-coupled ultralight scalar dark
  matter}.
\newblock 6 2020.

\bibitem{Lyth_1992}
D.~H. Lyth.
\newblock Axions and inflation: Vacuum fluctuations.
\newblock {\em Phys. Rev. D}, 45:3394--3404, May 1992.

\bibitem{Arvanitaki_anharmonic}
Asimina Arvanitaki, Savas Dimopoulos, Marios Galanis, Luis Lehner, Jedidiah~O.
  Thompson, and Ken Van~Tilburg.
\newblock {Large-misalignment mechanism for the formation of compact axion
  structures: Signatures from the QCD axion to fuzzy dark matter}.
\newblock {\em Phys. Rev. D}, 101(8):083014, 2020.

\bibitem{Arvanitaki:2019rax}
Asimina Arvanitaki, Savas Dimopoulos, Marios Galanis, Luis Lehner, Jedidiah~O.
  Thompson, and Ken Van~Tilburg.
\newblock {Large-misalignment mechanism for the formation of compact axion
  structures: Signatures from the QCD axion to fuzzy dark matter}.
\newblock {\em Phys. Rev. D}, 101(8):083014, 2020.

\bibitem{Huang:2020etx}
Junwu Huang, Amalia Madden, Davide Racco, and Mario Reig.
\newblock {Maximal axion misalignment from a minimal model}.
\newblock 6 2020.

\bibitem{Visinelli:2009zm}
Luca Visinelli and Paolo Gondolo.
\newblock {Dark Matter Axions Revisited}.
\newblock {\em Phys. Rev. D}, 80:035024, 2009.

\bibitem{Laine_book}
Mikko Laine and Aleksi Vuorinen.
\newblock {\em {Basics of Thermal Field Theory}}, volume 925.
\newblock Springer, 2016.

\bibitem{Kapusta_book}
J.I. Kapusta and Charles Gale.
\newblock {\em {Finite-temperature field theory: Principles and applications}}.
\newblock Cambridge Monographs on Mathematical Physics. Cambridge University
  Press, 2011.

\bibitem{Arnold_pureQCD}
Peter Arnold and Chengxing Zhai.
\newblock Three-loop free energy for pure gauge qcd.
\newblock {\em Phys. Rev. D}, 50:7603--7623, Dec 1994.

\bibitem{Arnold_QCD}
Peter Arnold and Chengxing Zhai.
\newblock Three-loop free energy for high-temperature qed and qcd with
  fermions.
\newblock {\em Phys. Rev. D}, 51:1906--1918, Feb 1995.

\end{thebibliography}
\bibliographystyle{unsrt}
\end{document}